\documentclass[%
preprint,
nofootinbib,
 amsmath,amssymb,
 aps,
 11pt,
]{revtex4}

\usepackage{multirow}
\usepackage{booktabs}
\usepackage{cancel}
\usepackage{color}
\usepackage{graphicx,physics,slashed}
\usepackage{graphicx}
\usepackage{hyperref}
\hypersetup{colorlinks=true, citecolor=blue, urlcolor=blue, linkcolor=blue}
\usepackage{dcolumn}
\usepackage{bm}
\usepackage{subfigure}
\usepackage{slashed}
\usepackage{setspace} 
\usepackage{ulem}
\usepackage{soul} 
\usepackage{cancel} 

\usepackage{float}
\usepackage{rotating}

\begin{document}

\title{\Large Transition form factors of the $\Lambda_b \rightarrow \Lambda(1520)$ in QCD light-cone sum rules }
\vspace{5mm}
 
\author{
Ke-Sheng Huang$^{1}$\footnote{huangksh21@lzu.edu.cn}
Hua-Yu Jiang$^2$\footnote{jianghuayu@htu.edu.cn, corresponding author}, 
Fu-Sheng Yu$^{1}$\footnote{yufsh@lzu.edu.cn,corresponding author}
}

\address{
$^1$Frontiers Science Center for Rare Isotopes, and School of Nuclear Science and Technology, Lanzhou University, Lanzhou 730000, China \\
$^2$Institute of Particle and Nuclear Physics, Henan Normal University, Xinxiang 453007, China\\
 }

\date{\today}

\begin{abstract}

In this work, we investigate the transition form factors for $\Lambda_b\rightarrow{\Lambda(1520)}$ within the framework of light-cone sum rules (LCSR), using the light-cone distribution amplitudes (LCDAs) of the $\Lambda_b$-baryon. In the hadronic representation of the correlation function, we carefully select the appropriate Lorentz structures and isolate the contributions from both the $\Lambda(1520)(J^P=(3/2)^-)$ and the $\Lambda(1890)(J^P=(3/2)^+)$, ensuring that the form factors for $\Lambda_b\rightarrow{\Lambda(1520)}$ can be calculated unambiguously. We also provide predictions for various physical observables in the decay $\Lambda_b\rightarrow{\Lambda(1520)}l^+l^-$, including the differential branching fraction, the lepton-side forward-backward asymmetry, the longitudinal polarization fraction, and the CP-averaged normalized angular observable. Our prediction for the differential branching fraction of $\Lambda_b\rightarrow{\Lambda(1520)}\mu^+\mu^-$ is in good agreement with the LHCb measurement within the uncertainties.

\end{abstract} 	


\maketitle
\vspace{10mm}

\section{Introduction}
The bottom hadron decays induced by flavor-changing neutral currents (FCNC), involving quark-level transitions such as $ b \rightarrow sl^{+}l^{-} $ or $ b \rightarrow s\gamma$, are prohibited at tree level within the Standard Model (SM) and can only take place through significantly suppressed loop diagrams.
The relevant observables are sensitive to the intermediate particles (in SM or beyond) in the loop.
As a result, the study of FCNC processes for bottom-meson and bottom-baryon decays is very important for checking if the Standard Model is correct and for finding hints of new physics (NP).

In the past decade, there have been substantial advancements in both the theoretical and experimental studies of FCNC processes involving $B$ mesons. 
These progresses have significantly improved our understandings of the underlying mechanisms, providing valuable insights into particle physics beyond the Standard Model. 
Theoretical models have become more refined, incorporating new computational techniques and precision calculations, while experimental efforts have led to more accurate measurements and new observations of rare decays. 
These progresses are paving the way for deeper exploration of phenomena like CP violation and new physics scenarios.
In particular, the transition form factors like $B \rightarrow K^{(*)}$ and $B \rightarrow \phi$ have been carefully calculated using Lattice QCD (LQCD) in the large momentum transfer region \cite{Horgan:2013hoa,Bailey:2015dka,Horgan:2015vla,Parrott:2022zte}. In addition to LQCD, other non-perturbative approaches, such as light-cone sum rules (LCSR), have also been applied to obtain these form factors \cite{Bharucha:2015bzk,Gubernari:2018wyi}. These combined efforts have led to more accurate and reliable predictions for these important transitions in $B$ meson physics.
Experimentally, a broad scope of physical observables, such as differential branching fractions and angular distributions, have been measured, providing valuable data to test theoretical predictions and refine our understanding of $B$-meson rare decays \cite{LHCb:2020gog,CMS:2020oqb,LHCb:2020lmf,LHCb:2021xxq}.
Until now, there are still some tensions between the SM predictions and the measurements for these physical observables, especially for the differential branching fractions in the large recoil regions in $B^+\to K^+\mu^+\mu^-$, $B^0\to K^0\mu^+\mu^-$, $B^+\to K^{*+}\mu^+\mu^-$ \cite{LHCb:2014cxe}, $B^0\to K^{*0}\mu^+\mu^-$ \cite{LHCb:2016ykl}, $B_s^0\to \phi\mu^+\mu^-$ \cite{LHCb:2021zwz}.
Furthermore, the tensions are systematically deviated since the theoretical predictions are larger than the experimental data in all of the above processes. 
It has to be extensively studied in other processes to clarify this anomaly as signal of new physics or not. 

Unlike $B$ meson decays, the $\Lambda_b$ baryon features a non-zero spin structure, which not only differentiates its decay behavior but also significantly enriches the spectrum of physical observables linked to FCNC processes in its decays. 
This added complexity provides a more nuanced framework for testing theoretical models and probing the dynamics of flavor-changing neutral currents (FCNCs) in baryonic systems.
From a theoretical perspective, the rare decay $\Lambda_b \rightarrow \Lambda l^+l^-$ has been investigated using a variety of methods, including LCSR \cite{Wang:2008sm,Wang:2009hra,Wang:2015ndk}, LQCD \cite{Detmold:2012vy,Detmold:2016pkz}, and dispersive analyses \cite{Blake:2022vfl}. 
On the experimental side, the LHCb Collaboration has measured the differential branching fraction and angular observables of $\Lambda_b \rightarrow \Lambda \mu^+\mu^-$ \cite{LHCb:2013uqx,LHCb:2015tgy,LHCb:2018jna}, along with CP violation \cite{LHCb:2017slr} and lepton flavor universality (LFU) \cite{LHCb:2019efc}. Similar to $B$ meson decays, the experimental data for the $\Lambda_b \rightarrow \Lambda \mu^+\mu^-$ process also show discrepancies with the Standard Model predictions.

The $\Lambda_b$ baryon can decay not only into the ground state $\Lambda$, but also into the excited $\Lambda(1520)$ state which can be easily reconstructed in $\Lambda_b\to \Lambda(1520)\mu^+\mu^-\to pK^-\mu^+\mu^-$ by LHCb with all the charged particles in the final states \cite{LHCb:2023ptw}. 
The $\Lambda(1520)$ has spin-parity $J^P = (3/2)^-$, in contrast to the ground state $\Lambda$ ($J^P = (1/2)^+$), providing more opportunities to probe new physics in the $b \rightarrow s l^+ l^-$ transitions.
However, there are few theoretical studies on the $\Lambda_b \rightarrow \Lambda(1520)$ transition. The transition form factors have been computed using the Non-relativistic Quark Model (NRQM) \cite{Mott:2011cx, Mott:2015zma}, Lattice QCD (LQCD) \cite{Meinel:2020owd, Meinel:2021mdj}, and the Light-front Quark Model (LFQM) \cite{Li:2022nim}. A dispersive analysis \cite{Amhis:2022vcd}, combined with LQCD results, has also been applied to this transition. Angular analyses of the $\Lambda_b \rightarrow \Lambda(1520)(\rightarrow N\bar{K}) l^+ l^-$ decay have been conducted in Refs. \cite{Descotes-Genon:2019dbw, Das:2020cpv} for massless and massive leptons, respectively.
Experimentally, the LHCb Collaboration \cite{LHCb:2023ptw} has measured the differential branching fraction of $\Lambda_b \rightarrow \Lambda(1520)\mu^+\mu^-$ for the first time. This branching fraction is highly sensitive to the relevant transition form factors. In the high-$q^2$ region ($q^2 > 15$ GeV$^2$), predictions from NRQM \cite{Mott:2011cx, Mott:2015zma}, LQCD \cite{Meinel:2020owd, Meinel:2021mdj}, LFQM \cite{Li:2022nim}, and dispersive analysis \cite{Amhis:2022vcd} are consistent with experimental measurements, showing minimal model dependence. However, these models diverge significantly in the low-$q^2$ region. In particular, predictions from LFQM \cite{Li:2022nim} and the dispersive analysis \cite{Amhis:2022vcd} exceed measurements by almost an order of magnitude, while NRQM \cite{Mott:2011cx, Mott:2015zma} predicts the opposite trend. This suggests that the low-$q^2$ differential branching fraction is particularly sensitive to the form factors in this range.
The form factors derived from LQCD \cite{Meinel:2020owd, Meinel:2021mdj} are only reliable in the high-$q^2$ region ($q^2 > 16$ GeV$^2$), leaving a gap in the low-$q^2$ region. Therefore, accurate calculations in the low-$q^2$ region are essential. The Light-Cone Sum Rules (LCSR), which are well suited for heavy-to-light form factors in the large recoil (low-$q^2$) region, have been successful in analyzing processes like the FCNC in $B$ meson decays and $\Lambda_b \rightarrow \Lambda$ decays. Therefore, it is valuable to extend the LCSR to the $\Lambda_b \rightarrow \Lambda(1520)$ transition.

In this work, we employ the LCSR to calculate the transition form factors of $\Lambda_b \to \Lambda(1520)$, derived from the $\Lambda_b$-baryon-to-vacuum correlation function. The light-cone distribution amplitudes of the $\Lambda_b$ baryon are incorporated into the sum rules, and the interpolating current for the $\Lambda(1520)$ is used. A key challenge in applying LCSR to the $\Lambda_b(\frac{1}{2}^+) \to \Lambda(1520)(\frac{3}{2}^-)$ decay is that the $J=\frac{3}{2}$ baryon interpolating current couples not only to the $J^P=\frac{3}{2}^{\pm}$ states but also to  the $J^P=\frac{1}{2}^{\pm}$ states, leading to contamination. To address this issue, we adopt the solution proposed in Ref. \cite{Aliev:2023tpk}, where specific Lorentz structures are chosen to suppress such contamination. Moreover, to avoid any ambiguity in the calculation of the form factors, we also include the contribution of the positive-parity $\Lambda^*(\frac{3}{2}^+)$ state in the hadronic representation of the correlation function. This ensures that no redundant Lorentz structures remain in the partonic representation of the correlation function. 
We provide the first results of the form factors of $\Lambda_b\to \Lambda(1520)$ from the QCD-inspired method in the low $q^2$ region. The results are consistent with the experimental data within the uncertainties.

This paper is organized as follows. In Section \ref{Framework}, we begin with the definition of the form factors, along with presenting the endpoint relations that exist between them; then the sum rules for the $\Lambda_b \rightarrow \Lambda(1520)$ form factors are displayed. The details of the numerical analysis for the form factors are collected in Section \ref{Results}, and we also give our predictions on some physical observables for $\Lambda_b \rightarrow \Lambda(1520)l^+l^-$. The last Section \ref{Summary} is reserved for the summary. The paper also contains some appendices, where the LCSR expressions for the form factors (App.\ref{sec:Appendix A}-\ref{sec:Appendix C}) and the correlation coefficient matrix for the fitting parameters are collected (App.\ref{sec:Appendix-D}).
\section{The LCSR of $\Lambda_{b}\rightarrow\Lambda(1520)$ form factors}
\label{Framework}
\subsection{$\Lambda_{b}\rightarrow \Lambda(1520)$ form factors}
The decay $\Lambda_b \to \Lambda(1520) l^+ l^-$ proceeds via the flavor-changing neutral current (FCNC) $b \to s l^+ l^-$ transition at the quark level, described by the corresponding effective Hamiltonian \cite{Das:2018iap,Liu:2019igt}:
\begin{align}
\label{equ::Hamiltonian}
    \mathcal{H}_{eff}(b\rightarrow s l^+l^-)=&-\frac{4G_F}{\sqrt2}V_{tb}V^*_{ts}\frac{\alpha_e}{4\pi}
    \Big\{C^{eff}_9(\mu,q^2)\bar{s}\gamma_{\mu}P_Lb(\bar{l}\gamma^{\mu}l)-\frac{2m_b}{q^2}C^{eff}_7(\mu)\bar{s}i\sigma_{\mu\nu}q^{\nu}P_Rb(\bar{l}\gamma^{\mu}l) \nonumber \\ 
    &+C_{10}(\mu)(\bar{s}\gamma_{\mu}P_Lb)(\bar{l}\gamma^{\mu}\gamma_5 l)\Big\}\ ,
\end{align} 
where $P_{L(R)}=(1\mp \gamma_{5})/2$ and $\sigma_{\mu\nu}=i[\gamma_{\mu},\gamma_{\nu}]/2$, 
$G_{F}=1.16637\times 10^{-5}\ \text{GeV}^{-2}$ is the Fermi coupling constant, $\alpha_e=1/137$ is the electromagnetic coupling constant and $|V_{tb}V^{*}_{ts}|=0.04117$ \cite{UTfit:2022hsi} is the product of the Cabibbo-Kobayashi-Maskqawa matrix elements. 
The effective Wilson coefficients $C^{eff}_{7}(\mu),\, C^{eff}_{9}(\mu,q^{2}),\,C_{10}(\mu)$, and the bottom quark mass $m_b$ are provided in Refs. \cite{Yan:2000dc, Azizi:2012vy, Li:2004vh, Buras:1994dj, Ali:1994bf}. 
To obtain the amplitude for the decay $\Lambda_b \to \Lambda(1520) l^+ l^- $, we need to evaluate the hadronic matrix elements:  
$\langle \Lambda(1520)(p,s')|\bar{s}\gamma_{\mu}(1-\gamma_{5})b|\Lambda_b(p+q,s)\rangle$ and $\langle \Lambda(1520)(p,s')|\bar{s}i\sigma_{\mu\nu}q^{\nu}(1+\gamma_{5})b|\Lambda_b(p+q,s)\rangle$. 
For the rest of this work, we refer to $\Lambda(1520)$ as $\Lambda^{*}_{-}$. These matrix elements are parameterized in terms of fourteen independent form factors \cite{Mott:2011cx, Descotes-Genon:2019dbw, Das:2020cpv}. We adopt the helicity-based approach \cite{Descotes-Genon:2019dbw, Das:2020cpv}, which is particularly useful for calculating the helicity amplitudes.
\begin{align}
\label{FF:vector}
&\langle \Lambda^*_{-}(p,s')| \bar{s}\gamma_{\mu}b|\Lambda_{b}(p+q,s)\rangle\nonumber \\
=&\bar{u}^{\alpha}(p,s') \Bigg\{(p+q)_{\alpha} \bigg [ f^V_{t}(q^2)(m_{\Lambda_b}-m_{\Lambda^*_{-}})\frac{q_{\mu}}{q^2} \nonumber \\
&+f^V_{0}(q^2)\frac{m_{\Lambda_b} +m_{\Lambda^*_{-}}}{s_{+}} \left((q+p)_{\mu}+p_{\mu}-(m^2_{\Lambda_{b}}-m^2_{\Lambda^*_{-}})\frac{q_{\mu}}{q^2}\right) \nonumber\\
&+f^{V}_{\perp}(q^2)\left(\gamma_{\mu}-\frac{2m_{\Lambda^*_{-}}}{s_{+}}(p+q)_{\mu}-\frac{2m_{\Lambda_b}}{s_{+}}p_{\mu}\right) \bigg ] \nonumber \\
&+f^V_{g}(q^2)\bigg[g_{\alpha\mu}+m_{\Lambda^*_-}\frac{(p+q)_{\alpha}}{s_-}\bigg(\gamma_{\mu}-\frac{2p_{\mu}}{m_{\Lambda^*_-}}+\frac{2(m_{\Lambda^*_-}(p+q)_{\mu}+m_{\Lambda_b}p_{\mu})}{s_+}\bigg)\bigg]\Bigg\}u_{\Lambda_b}(p+q,s)\ , \\
&\langle \Lambda^*_{-}(p,s')| \bar{s}\gamma_{\mu}\gamma_{5}b|\Lambda_{b}(p+q,s)\rangle\nonumber\\
=&-\bar{u}^{\alpha}(p,s') \gamma_5\Bigg\{(p+q)_{\alpha} \bigg [ g^A_{t}(q^2)(m_{\Lambda_b}+m_{\Lambda^*_{-}})\frac{q_{\mu}}{q^2} \nonumber \\
&+g^A_{0}(q^2)\frac{m_{\Lambda_b} -m_{\Lambda^*_{-}}}{s_{-}} \left((q+p)_{\mu}+p_{\mu}-(m^2_{\Lambda_{b}}-m^2_{\Lambda^*_{-}})\frac{q_{\mu}}{q^2}\right) \nonumber\\
&+g^{A}_{\perp}(q^2)\left(\gamma_{\mu}+\frac{2m_{\Lambda^*_{-}}}{s_{-}}(p+q)_{\mu}-\frac{2m_{\Lambda_b}}{s_{-}}p_{\mu}\right) \bigg ] \nonumber \\
&+g^A_{g}(q^2)\bigg[g_{\alpha\mu}-m_{\Lambda^*_{-}}\frac{(p+q)_{\alpha}}{s_+}\bigg(\gamma_{\mu}+\frac{2p_{\mu}}{m_{\Lambda^*_{-}}}-\frac{2(m_{\Lambda^*_{-}}(p+q)_{\mu}-m_{\Lambda_b}p_{\mu})}{s_-}\bigg)\bigg]\Bigg\}u_{\Lambda_b}(p+q,s)\ , \label{FF:axial}
 \\
&\langle \Lambda^*_{-}(p,s')| \bar{s}i\sigma_{\mu\nu}q^{\nu}b|\Lambda_{b}(p+q,s)\rangle\nonumber\\
=&-\bar{u}^{\alpha}(p,s') \Bigg\{(p+q)_{\alpha} \bigg [ f^T_{0}(q^2)\frac{q^2}{s_+}\left((q+p)_{\mu}+p_{\mu}-(m^2_{\Lambda_{b}}-m^2_{\Lambda^*_{-}})\frac{q_{\mu}}{q^2}\right) \nonumber \\
&+f^{T}_{\perp}(q^2)(m_{\Lambda_b}+m_{\Lambda^*_{-}})\left(\gamma_{\mu}-\frac{2m_{\Lambda^*_{-}}}{s_{+}}(p+q)_{\mu}-\frac{2m_{\Lambda_b}}{s_{+}}p_{\mu}\right)\bigg ] \nonumber \\
&+f^T_{g}(q^2)\bigg[g_{\alpha\mu}+m_{\Lambda^*_{-}}\frac{(p+q)_{\alpha}}{s_-}\bigg(\gamma_{\mu}-\frac{2p_{\mu}}{m_{\Lambda^*_{-}}}+\frac{2(m_{\Lambda^*_{-}}(p+q)_{\mu}+m_{\Lambda_b}p_{\mu})}{s_+}\bigg)\bigg]\Bigg\}u_{\Lambda_b}(p+q,s)\ ,\label{TFF:vector} \\
\label{TFF:axial}
&\langle \Lambda^*_{-}(p,s')| \bar{s}i\sigma_{\mu\nu}q^{\nu}\gamma_5b|\Lambda_{b}(p+q,s)\rangle \nonumber\\
=&-\bar{u}^{\alpha}(p,s')\gamma_5 \Bigg\{(p+q)_{\alpha} \bigg [ g^{T5}_{0}(q^2)\frac{q^2}{s_-}\left((q+p)_{\mu}+p_{\mu}-(m^2_{\Lambda_{b}}-m^2_{\Lambda^*_{-}})\frac{q_{\mu}}{q^2}\right) \nonumber \\
&+g^{T5}_{\perp}(q^2)(m_{\Lambda_b}-m_{\Lambda^*_{-}})\left(\gamma_{\mu}+\frac{2m_{\Lambda^*_{-}}}{s_{-}}(p+q)_{\mu}-\frac{2m_{\Lambda_b}}{s_{-}}p_{\mu}\right)\bigg ] \nonumber \\
&+g^{T5}_{g}(q^2)\bigg[g_{\alpha\mu}-m_{\Lambda^*_{-}}\frac{(p+q)_{\alpha}}{s_+}\bigg(\gamma_{\mu}+\frac{2p_{\mu}}{m_{\Lambda^*_{-}}}-\frac{2(m_{\Lambda^*_{-}}(p+q)_{\mu}-m_{\Lambda_b}p_{\mu})}{s_-}\bigg)\bigg]\Bigg\}u_{\Lambda_b}(p+q,s)\ ,
\end{align}
where $u^{\alpha}(p,s')$ and $u_{\Lambda_b}(p+q,s)$ are the Rarita-Schwinger (RS) and Dirac spinors for the $\Lambda^{\ast}_{-}$ and $\Lambda_b$ baryons, respectively, where $\{m_{\Lambda_b}, p, s'\}$ and $\{m_{\Lambda^{\ast}_{-}}, p+q, s\}$ represent the mass, momentum, and spin of the $\Lambda_b$ and $\Lambda^{\ast}_{-}$ baryons. The form factors $f^{\Gamma}_{i}(q^2)$ and $g^{\Gamma}_{i}(q^2)$ depend on the momentum transfer $q^2$, and the kinematic variable $s_{\pm} = (m_{\Lambda_b} \pm m_{\Lambda^{\ast}_{-}})^2 - q^2$.

Although the fourteen form factors $f^{\Gamma}_{i}(q^2)$ and $g^{\Gamma}_{i}(q^2)$ are defined independently in principle, certain relations between them emerge at specific values of $q^2$. These relations, known as endpoint relations, arise from two distinct mechanisms.
First, the hadronic matrix elements $\langle \Lambda^{\ast}_{-}(p,s')|\bar{s}\Gamma b|\Lambda_{b}(p+q,s)\rangle$ must be free of kinematic singularities, which occur at $q^2 = 0$ and $q^2 = q^2_{max} \equiv (m_{\Lambda_b} - m_{\Lambda^{\ast}_{-}})^2 $. These singularities can be removed using the following identities:

\begin{eqnarray}
\label{equ;;end relations1}
& f^{V}_{t}(0)=f^{V}_{0}(0)\, , \ \ \ \ g^{A}_{t}(0)=g^{A}_{0}(0)\, ,\ \ \ \ f^{V}_{g}(q^2_{max}) =f^{T}_{g}(q^2_{max})=0\, ,  \nonumber\\
& g^{A}_{0}(q^2_{max})=g^{A}_{\perp}(q^2_{max})+\frac{1}{4m_{\Lambda_b}}g^{A}_{g}(q^2_{max}) \,,\nonumber \\ 
& g^{T5}_{0}(q^2_{max})=g^{T5}_{\perp}(q^2_{max})+\frac{1}{4m_{\Lambda_b}(m_{\Lambda_b}- m_{\Lambda^*_{-} })}g^{T5}_{g}(q^2_{max})\,.
\end{eqnarray}
In addition to the identities above, the algebraic relation between $\sigma^{\mu\nu}$ and $\sigma^{\mu\nu}\gamma_{5}$ ensures that: 
\begin{eqnarray}
\label{equ;;end relations2}
  g^{T5}_{\perp}(0)= f^{T}_{\perp}(0)\,, \ \ \ \  g^{T5}_{g}(0)= f^{T}_{g}(0)\frac{(m_{\Lambda_b}+m_{\Lambda^*_{-}})}{(m_{\Lambda_b}-m_{\Lambda^*_{-}})}\,.
\end{eqnarray}
Further discussions on the endpoint relations for the baryon form factors can be found in Ref. \cite{Hiller:2021zth}. In the following, we will apply these endpoint relations Eqs.(\ref{equ;;end relations1})–(\ref{equ;;end relations2}) to extrapolate the LCSR computations of the form factors $f^{\Gamma}_{i}(q^2)$ and $g^{\Gamma}_{i}(q^2)$ across the entire physical kinematic region.

\subsection{Interpolating current and correlation function}
To calculate the form factors governing the transition $\Lambda_{b}(\frac{1}{2}^{+})\rightarrow \Lambda^{*}_{-}(\frac{3}{2}^{-})$, we begin with the following two-point correlation function, which is sandwiched between the vacuum and the on-shell $\Lambda_{b}$-baryon state:
\begin{align} 
\label{equ;;correlation function}
    \Pi_{\lambda\mu}(p,q)=i\int d^4 x e^{ipx}\langle 0
|\mathcal{T}\{\eta_{\lambda}(x),j_{\mu,d}(0)|\Lambda_b(p+q)\rangle \ ,
\end{align}
here, the local current $\eta_{\lambda}$ is the interpolating current for the $\Lambda^{*}_{-}$ baryon, while $j_{\mu,d}$ represents the heavy-light weak transition current $\bar{s}\Gamma_{\mu,d}b$, with the index "$d$" denoting a specific Lorentz structure, i.e.,
\begin{eqnarray}
\label{equ;;Lorentz structure}
  &j_{\mu,V}=\bar{s}\gamma_{\mu}b, \ \ \ \ j_{\mu,A}=\bar{s}\gamma_{\mu}\gamma_5 b, \nonumber  \\
     &j_{\mu,T}=\bar{s}i\sigma_{\mu\nu}q^{\nu}b,\ \ \ \  j_{\mu,T5}=\bar{s}i\sigma_{\mu\nu}q^{\nu}\gamma_5 b, 
\end{eqnarray}
The interpolating current for the $\Lambda^*_{-}$ baryon is chosen with the following formula from the Ref.\cite{Lee:2002jb}:
\begin{align}
\label{equ;;interpolating current}
    \eta_{\lambda}=\sqrt{\frac{1}{6}}\varepsilon^{abc}[2(u^{aT}C\sigma_{k\delta}d^{b})\sigma^{k\delta}\gamma_{\lambda}s^{c}+(u^{aT}C\sigma_{k\delta}s^{b})\sigma^{k\delta}\gamma_{\lambda}d^{c}-(d^{aT}C\sigma_{k\delta}s^{b})\sigma^{k\delta}\gamma_{\lambda}u^{c}],
\end{align}
where $T$ represents the transpose matrix, $C$ is the charge conjugation matrix, and the sum runs over the color indices $ a, b, c$. It should be noted that the interpolating current $\eta_{\lambda}(x)$ not only couples to the negative-parity $J^P=\frac{3}{2}^{-}$ state, but also couples to the positive-parity $J^P=\frac{3}{2}^+$ state. 
Consequently, the hadronic representation of the correlation function receives an additional contribution from the positive-parity resonance $\Lambda(1890)(\frac{3}{2}^{+})$. 
This implies that the contribution from the $\Lambda(1890)(\frac{3}{2}^{+})$ state must be taken into account in our calculation.
For the final state $\Lambda(1890)(\frac{3}{2}^{+})$, the form factors for the transition $\Lambda_{b}\rightarrow \Lambda(1890)(\frac{3}{2}^{+})$ are similar to those for $\Lambda_{b}\rightarrow \Lambda(1520)(\frac{3}{2}^{-})$, except for an additional $\gamma_{5}$ factor on the right-hand side of the Rarita-Schwinger spinors $\bar{u}^{\alpha}(p,s')$. Additionally, the Lorentz structure of the form factors $f^{V}(q^2)$ and $f^{T}(q^2)$ is modified to that of $g^{A}(q^2)$ and $g^{T5}(q^2)$ to account for the parity transformation.

Following the standard strategy of Light-Cone Sum Rules (LCSR), the correlation function is treated both in the hadronic and partonic representations. The results from these two representations are then matched using the dispersion relation. In both representations, the correlation function can be decomposed into eight distinct Lorentz structures corresponding to each weak transition current $j_{\mu,d}$. By matching the coefficients of these Lorentz structures in the two representations, one can isolate and distinguish the contributions from the $\Lambda(1520)$ baryon with $J^{P}=\frac{3}{2}^{-}$ and the $\Lambda(1890)$ baryon with $J^{P}=\frac{3}{2}^{+}$ by solving a system of linear equations. 
Next, the quark-hadron duality assumption and Borel transformation are applied to subtract and suppress the contributions from higher excited states and the continuum.

In the hadronic representation, one can insert a complete set of states with the same quantum number as the interpolating current $\eta_{\lambda}$ and isolate the contributions of the state $\Lambda(1520)(\frac{3}{2}^{-})$ and its positive partner $\Lambda(1890)(\frac{3}{2}^{+})$. Then, the correlation function can be expressed in terms of the hadronic matrix elements: 
\begin{align}
\label{equ::hadronic level}
\Pi_{\lambda\mu}=&\frac{1}{m^2_{\Lambda^{*}_-}- p^2}\sum_{s'}\langle 0|\eta_{\lambda}|\Lambda^{*}_{-}(p,s')\rangle\langle\Lambda^{*}_-(p,s') | j_{\mu,a}|\Lambda_b(p+q,s)\rangle \nonumber \\
&+\frac{1}{m^2_{\Lambda^{*}_+}- p^2}\sum_{s'}\langle 0|\eta_{\lambda}|\Lambda^{*}_{+}(p,s')\rangle\langle\Lambda^{*}_+(p,s') | j_{\mu,a}|\Lambda_b(p+q,s)\rangle + ... \, ,
\end{align}
where the ellipsis stands for the contributions from the higher excited and continuum states. The $\Lambda^*_{-(+)}$ is the negative(positive)-parity baryon $\Lambda(1520)(\Lambda(1890))$. The coupling constants of the $\Lambda^*_{\pm}$ baryons for the interpolating current $\eta_{\lambda}$ are defined as following:
\begin{align}
\label{equ::decay constant}
\langle 0|\eta_{\lambda}|\Lambda^{*}_{+}(p,s')&=\lambda_{+}u_{\lambda}(p,s') \, , \nonumber  \\ 
\langle 0|\eta_{\lambda}|\Lambda^{*}_{-}(p,s')\rangle &=\lambda_{-}\gamma_5 u_{\lambda}(p,s') \, ,
\end{align}
Using the definition of the form factors and the coupling constants and performing the spin-summation of the Rarita-Schwinger spinors:
\begin{align}
\label{equ::RSsum}
\sum_{s'} u_{\lambda}(p,s')\bar{u}_{\alpha}(p)=-(\not p+m_{\Lambda^{*}_{\pm}})\Big[g_{\lambda\alpha}-\frac{1}{3}\gamma_{\lambda}\gamma_{\alpha}-\frac{2}{3}\frac{p_{\lambda}p_{\alpha}}{m^2_{\Lambda^{*}_{\pm}}}+\frac{1}{3}\frac{p_{\lambda}\gamma_{\alpha}-p_{\alpha}\gamma_{\lambda}}{m_{\Lambda^{*}_{\pm}}}\Big]\ ,
\end{align}
we can simplify the correlation function in the  hadronic representation.
Before proceeding to the next step, it is necessary to make some preliminary remarks:
\begin{itemize}
    \item The interpolating current $\eta_{\lambda}$ for the spin $\frac{3}{2}$ baryon has nonzero overlap not only with spin $\frac{3}{2}$ baryons, but also with spin $\frac{1}{2}$ baryons. In fact, the coupling constants of the spin $\frac{1}{2}^{\pm}$ baryons with the interpolating current $\eta_{\lambda}$ are defined as:
    \begin{align}
\langle 0|\eta_{\lambda}|\Lambda(\frac{1}{2}^{\pm})(p,s')\rangle \sim(\gamma_{\lambda}+\frac{4p_{\lambda}}{m_{\pm}})(\gamma_5) u(p,s')\ ,
\end{align}
    we can observe that this expression contains the Lorentz structures $p_{\lambda}$ or $\gamma_{\lambda}$, indicating that these structures include contributions from spin-$\frac{1}{2}$ baryons. From Eq. (\ref{equ::RSsum}), we see that the Lorentz structure $g_{\lambda\alpha}$ can only receive contributions from spin-$\frac{3}{2}$ baryons, meaning there is no contamination from spin-$\frac{1}{2}$ baryons. Therefore, we can focus exclusively on the contribution from the structure $g_{\lambda\alpha}$ in the spin summation over the Rarita-Schwinger spinors, which effectively excludes any contributions from spin-$\frac{1}{2}$ baryons.
    \item From Eqs. (\ref{equ::decay constant}) and (\ref{equ::RSsum}), we can see that multiple Lorentz structures contribute to the hadronic representation of the correlation function. However, not all of these structures are necessary. By applying the equation of motion, we are left with only the following form of the correlation function in the hadronic representation, which is decomposed into eight independent and distinct Lorentz structures:
\begin{align}
\label{equ::Hadronic decomposed}
&\Pi_{\lambda\mu,d}(p,q)= 
\Big[\Pi^d_1 g_{\lambda\mu}+\Pi^d_2g_{\lambda\mu}\not q \nonumber\\
&+q_{\lambda}(\Pi^d_3p_{\mu}+\Pi^d_4 p_{\mu}\not q+\Pi^d_5\gamma_{\mu}+\Pi^d_{6}\gamma_{\mu}\not q+\Pi^d_{7}q_{\mu}+\Pi^d_{8}q_{\mu}\not q)\Big]u_{\Lambda_b}(p+q)\ ,
\end{align}
  Furthermore, one might wonder whether the Lorentz structure $\gamma_{\lambda}\gamma_{\mu} \not{q} $ would contribute, since the following relation holds:
   \begin{align}
   \label{equ::indepent Eq}
\gamma_{\lambda}\gamma_{\mu}\not q=g_{\lambda\mu}\not q -q_{\lambda}\gamma_{\mu} +\gamma_{\lambda}q_{\mu}-i\epsilon_{\nu\lambda\mu\beta}\gamma^{\nu}q^{\beta}\gamma_5\ .
\end{align}
 In fact, we can subtract these contributions when deriving the final expression in the hadronic representation of the correlation function, due to the identical coefficients multiplying each term on the right-hand side of Eq. (\ref{equ::indepent Eq}). These techniques can also be applied to the partonic representation.
\end{itemize}

With these considerations in mind, we can derive the hadronic expression for the correlation function. The detailed expressions in the hadronic representation are summarized in Appendix \ref{sec:Appendix A}.

\subsection{The light-cone sum rules for the form factor}
After obtaining the correlation function in the hadronic representation, we now turn our attention to the partonic representation by applying the light-cone operator product expansion. By using Wick's theorem on the correlation function, we can derive the results at the quark-gluon level:
\begin{align}
\label{equ::partonic side}
&\Pi_{\lambda\mu,d}(p,q)=i\int d^4 x e^{ixp}\ \sqrt{\frac{1}{6}}\varepsilon^{abc}\nonumber \\
&\times\Bigg [2(C\sigma_{k\delta})_{\alpha\beta}(\sigma^{k\delta}\gamma_{\lambda})_{\gamma'\gamma}[\Gamma_{\mu,d}]_{\rho\tau}S(x)_{\gamma\rho}\times\langle 0 |u^{aT}(x)_{\alpha}d^{b}(x)_{\beta}b(0)^{c}_{\tau}|\Lambda_b(p+q)\rangle \nonumber \\
&+(C\sigma_{k\delta})_{\alpha\beta}(\sigma^{k\delta}\gamma_{\lambda})_{\gamma'\gamma}[\Gamma_{\mu,d}]_{\rho\tau}S(x)_{\beta\rho}\times\langle 0 |u^{aT}(x)_{\alpha}d^{b}(x)_{\gamma}b(0)^{c}_{\tau}|\Lambda_b(p+q)\rangle \nonumber \\
&-(C\sigma_{k\delta})_{\alpha\beta}(\sigma^{k\delta}\gamma_{\lambda})_{\gamma'\gamma}[\Gamma_{\mu,d}]_{\rho\tau}S(x)_{\beta\rho}\times\langle 0 |d^{aT}(x)_{\alpha}u^{b}(x)_{\gamma}b(0)^{c}_{\tau}|\Lambda_b(p+q)\rangle\Bigg]\ ,
\end{align}
where the $S(x)$ represents the free propagator of the strange quark. 
The matrix element $\varepsilon^{abc}\langle 0 |u(x)^{aT}_{\alpha}d(x)^{b}_{\beta}b(0)^{c}_{\tau}|\Lambda_b(v)\rangle$ defines the light-cone distribution amplitudes (LCDAs) of the $\Lambda_b$ baryon, which serve as the crucial non-perturbative input for the theoretical calculation of the correlation function in the partonic representation. This matrix element can be parameterized in terms of four distinct distribution amplitudes \cite{Ball:2008fw,Ali:2012pn,Bell:2013tfa}.
\begin{align}
\label{equ::LCDAs}
&\varepsilon^{abc}\langle 0 |u(t_{1}n)^{Ta}_{\alpha}d(t_{2}n)^{b}_{\beta}b(0)^{c}_{\tau}|\Lambda_{b}(v)\rangle \nonumber \\
&=\frac{1}{8}f^{(2)}_{\Lambda_{b}}\Psi_2(t_1,t_2)(\not \bar{n}\gamma_5 C^{-1})_{\beta\alpha}u_{\Lambda_b}(v)_{\tau}+\frac{1}{4}f^{(1)}_{\Lambda_{b}}\Psi^{s}_3(t_1,t_2)(\gamma_5C^{-1})_{\beta\alpha}u_{\Lambda_b}(v)_{\tau} \nonumber \\
&-\frac{1}{8}f^{(1)}_{\Lambda_{b}}\Psi^{\sigma}_{3}(t_1,t_2)(i\sigma_{n\bar{n}}\gamma_5C^{-1})_{\beta\alpha}u_{\Lambda_b}(v)_{\tau}+\frac{1}{8}f^{(2)}_{\Lambda_{b}}\Psi_4(t_1,t_2)(\not n\gamma_5C^{-1})_{\beta\alpha}u_{\Lambda_b}(v)_{\tau}\ ,
\end{align}
where the $n_{\mu}=x_{\mu}/v \cdot x$ and $\bar{n}_{\mu}=2v_{\mu}-n_{\mu}$ represent two light-cone vectors ($v_{\mu}$ is the velocity of the $\Lambda_b$ baryon), satisfying the conditions $n^2=\bar{n}^2=0$ and $n_{\mu}\bar{n}^{\mu}=2$, $f^{(1)}_{\Lambda_b}$ and $f^{(2)}_{\Lambda_b}$ are the decay constants of the $\Lambda_{b}$ baryon and $\Psi_{i}(t_1,t_2)$ are the DAs of $\Lambda_{b}$ baryon with definite twist. 
The matrix element $\varepsilon^{abc}\langle 0 |d(x)^{aT}_{\alpha}u(x)^{b}_{\beta}b(0)^{c}_{\tau}|\Lambda_b(v)\rangle$ in the last line of Eq. (\ref{equ::partonic side}) is identical to the definition of the LCDAs of $\Lambda_b$ baryon, except for an additional minus sign due to the isospin-zero condition of the diquark.

In the above expression, Eq. (\ref{equ::LCDAs}), the LCDAs $\Psi_{i}(t_1,t_2)$ are given in position space, where $t_1$ and $t_2$ denote the positions of the light quarks within the diquark. In some cases, it is more convenient to transition to momentum space by performing a Fourier transform to facilitate the calculations.
\begin{align}
\label{equ::Fourier}
    \Psi_{i}(t_1,t_2)&=\int^{\infty}_0 d\omega_1\int d\omega_2 e^{-i\omega_1t_1-i\omega_2t_2}\psi_{i}(\omega_{1},\omega_{2}) \nonumber \\
    &=\int^{\infty}_0 \omega d\omega\int^{1}_{0} du e^{-i\omega(t_1u+it_2\bar{u})}\psi_{i}(\omega,u)\nonumber \\
    &=\int^{\infty}_0 \omega d\omega\int^{1}_{0} du e^{-i\omega v\cdot x} \psi_{i}(\omega,u) \, , \ \ (t_1=t_2=v\cdot x) \,,
\end{align}
here, $\omega_{1}$ and $\omega_2$ represent the energies of the up and down quark, respectively, while $\omega$ denotes the total energy carried by the light quarks, and $u$ is the energy fraction carried by the up quark. There are various models of LCDAs for the $\Lambda_{b}$ baryon in the literature \cite{Ball:2008fw, Ali:2012pn, Bell:2013tfa}. In this work, we adopt the exponential model motivated by the "on-shell wave function" analysis \cite{Bell:2013tfa}, which has been widely employed in studies of CP violation in $\Lambda_{b}$ decays \cite{Han:2022srw, Yu:2024cjd}. The relevant formulas are provided below:
\begin{eqnarray}
\label{equ::exp-model}
    \psi_2(\omega,u)&=&\frac{\omega^2 u(1-u)}{\omega^4_0}e^{-\omega/\omega_0} \, , \nonumber \\
    \psi^{s}_3(\omega, u)&=&\frac{\omega}{2\omega^3_0}e^{-\omega/\omega_0} \, ,\nonumber \\
    \psi^{\sigma}_3(\omega, u)&=&\frac{\omega (2u-1)}{2\omega^3_0}e^{-\omega/\omega_0} \, ,\nonumber \\
        \psi_4(\omega, u)&=&\frac{1}{\omega^2_0}e^{-\omega/\omega_0} \, ,
\end{eqnarray}
where $\omega_{0}=0.28\pm 0.05\ \text{GeV}$ \cite{Wang:2015ndk}. 

By substituting the LCDAs of the $\Lambda_b$ baryon, given by Eqs.(\ref{equ::LCDAs})-(\ref{equ::Fourier}), into the correlation function in the partonic representation Eq.(\ref{equ::partonic side}), and performing the necessary calculations and simplifications, we obtain a set of Lorentz structures that are analogous to those found in the hadronic representation Eq.(\ref{equ::Hadronic decomposed}). The coefficients of each Lorentz structure in the partonic representation can then be expressed as:
\begin{eqnarray}
\label{EQ:invaramp}
	\Pi^{d}_{i}(p^2 , q^2)=\sum_{n=1,2}\int^{\infty}_0d\omega\int^1_0du \frac{\rho^{d}_{i,n}(p^2, q^2)}{D^n} \, ,
\end{eqnarray}
with the denominator
\begin{eqnarray}
D=\bar{\sigma}p^2+\sigma q^2-\sigma\bar{\sigma}m^2_{\Lambda_b}-m^2_s \,,
\end{eqnarray}
where  $\sigma={\omega}/{m_{\Lambda_b}} ,\  \bar{\sigma}=1-\sigma$. 
The functions $\rho^d_{i,n}(p^2, q^2)$ are distinguished by their indices: $d = V, A, T, T5$ (which correspond to the Lorentz structures of the weak transition current $\bar{s} \Gamma_{\mu,d} b)$, $i = 1, \dots, 8$ (denoting the coefficients of each Lorentz structure in the partonic representation), and $n = 1, 2$ (indicating the power of the denominator).

The results in the hadronic and partonic representations are matched using the dispersion relation. To obtain the expressions for the form factors of $\Lambda_{b}\rightarrow \Lambda(1520)$ without ambiguity, we equate the coefficients of each Lorentz structure in both representations and solve the resulting system of linear equations to eliminate the contribution from the positive baryon $\Lambda(1890)$. The final expressions for the form factors of $\Lambda_{b}\rightarrow \Lambda(1520)$ are compiled in Appendix \ref{sec:Appendix C}. 
During this process, the quark-hadron duality and the Borel transformation are applied to subtract and suppress contributions from higher excited and continuum states. These three steps can be implemented through the following substitution:
\begin{eqnarray}
\label{equ::Borel transformation}
\int d\omega\int du\frac{\rho(p^2 , q^2)}{D}&\rightarrow& -m_{\Lambda_b} \int^1_0 du\int^{\sigma_0}_{0}\frac{d\sigma}{\bar{\sigma}}\rho(s,q^2)\exp\Big({-\frac{s}{M^2}}\Big) \, , \nonumber \\
\int d\omega\int du\frac{\rho(p^2 , q^2)}{D^2}&\rightarrow& \frac{m_{\Lambda_b}}{M^2} \int^1_0 du\int^{\sigma_0}_{0}\frac{d\sigma}{\bar{\sigma}^2}\rho(s,q^2)\exp\Big({-\frac{s}{M^2}}\Big)+\int^1_0 du\frac{\rho(s_0,q^2)\eta(\sigma_0)e^{-s_0/M^2}}{\bar{\sigma}^2_0m_{\Lambda_b}} \, , \nonumber \\
\end{eqnarray}
where the transformed coefficient functions, $\rho^{d}_{i,n}(s,q^2)$, are presented in Appendix.\ref{sec:Appendix B} and the involved parameters are defined as:
\begin{eqnarray}
s(\sigma)=\sigma m_{\Lambda_b}^2+\frac{m^2_s-\sigma q^2}{\bar \sigma},\,\,\, \eta(\sigma_0)=\left(1+\frac{m^2_s-q^2}{\bar \sigma^2_0m_{\Lambda_b}^2}\right)^{-1}\, ,
\end{eqnarray}
and $\sigma_0$ is the positive solution of the corresponding quadratic equation for $s=s_0$:
\begin{eqnarray}
     \sigma_0=\frac{(s_0+m^2_{\Lambda_b}-q^2)-\sqrt{(s_0+m^2_{\Lambda_b}-q^2)^2-4m^2_{\Lambda_b}(s_0-m^2_s)}}{2m^2_{\Lambda_b}}\ .
\end{eqnarray}
where $s_0$ is the effective threshold parameter.
\section{Numerical analysis}
\label{Results}
\subsection{The results for the form factors}
\begin{figure}
\centering
    \subfigure{\includegraphics[width=0.45\linewidth]{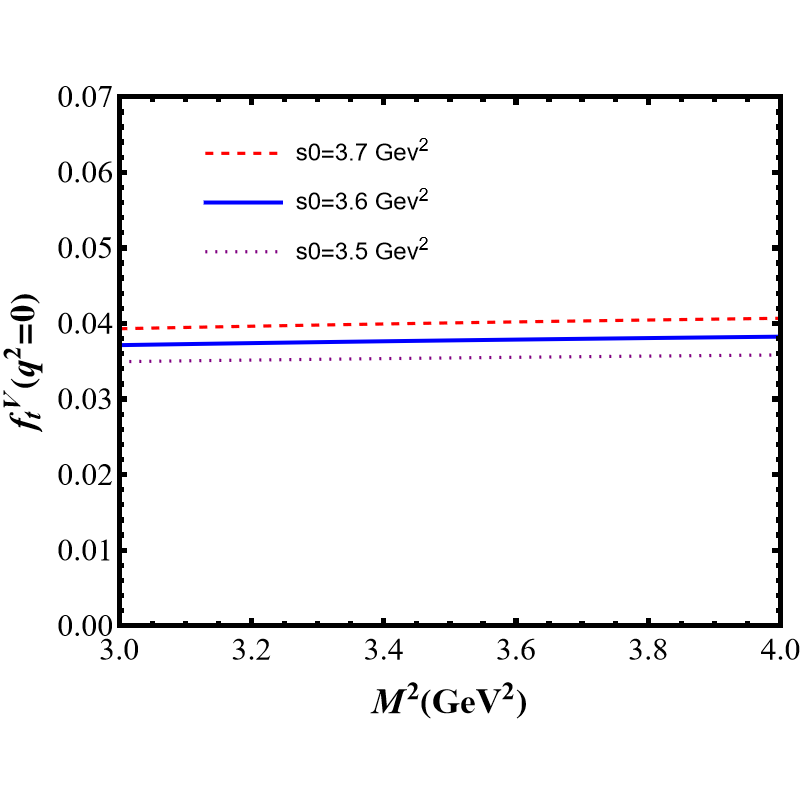}}
    \quad
     \subfigure{\includegraphics[width=0.45\linewidth]{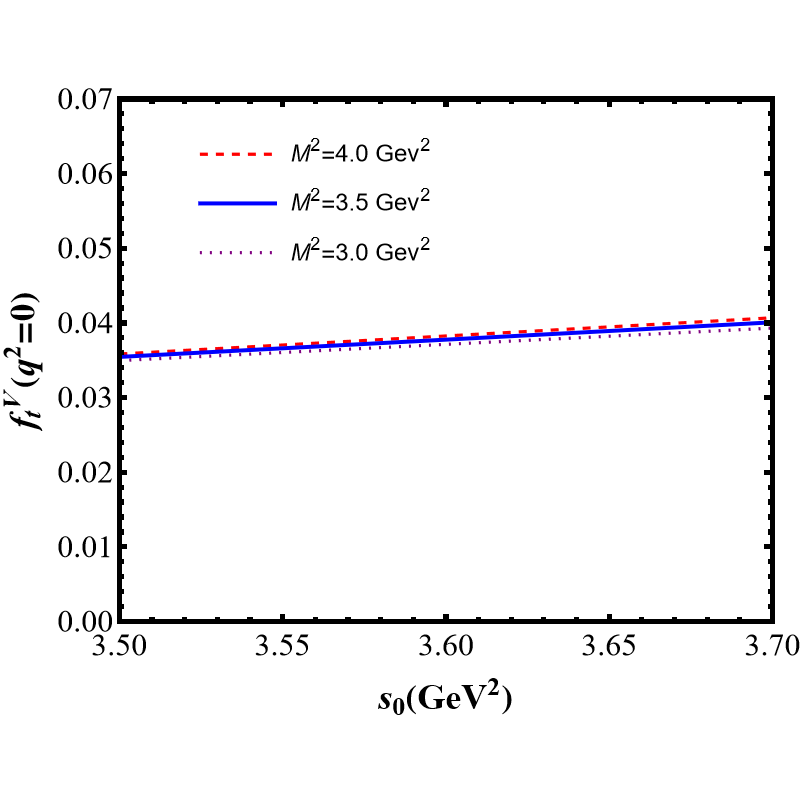}}
     \\
\caption{The dependence of $f^{V}_{t}(q^2=0)$ on the Borel parameter $M^2$ (Left) and the effective threshold parameters $s_0$ (Right). Dashed, solid and dotted curves are the numerical results of $f^{V}_{t}(q^2=0)$  corresponding to $s_0=3.7\text{GeV}^2,\ 3.6\text{GeV}^2,\ 3.6\text{GeV}^2$ and $M^2=4.0\text{GeV}^2,\ 3.5\text{GeV}^2,\ 3.0\text{GeV}^2$ respectively.}
\label{fig::dependence on M^2&s_0}
\end{figure}

This subsection is devoted to the numerical analysis of the sum rules for the $\Lambda_{b}\rightarrow \Lambda(1520)$ transition form factors, derived in the previous section. Before proceeding with the numerical analysis, it is important to first gather the essential input parameters for the LCSR. All the required input parameters for the LCSR predictions and the subsequent phenomenological analysis are listed in Table \ref{Tab::input parameters}.
 \begin{table}
  \caption{Numerical values of the theatrical input parameters employed in LCSR prediction of the  $\Lambda_{b}\rightarrow \Lambda(1520)$ form factors as well as the subsequent phenomenological analysis for the $\Lambda_{b}\rightarrow \Lambda(1520)l^{+}l^{-}$ physical observables.}
  \label{Tab::input parameters}
  \centering
     \begin{tabular}{|l |l r|}
     \hline
     \hline
     \text{Parameter}&\text{Value}&\text{Ref.}\\
     \hline
        $m_{\Lambda_{b}}$ & 5.620\ \text{GeV}  &  \cite{ParticleDataGroup:2024cfk} \\
        $m_{\Lambda^{*}_{-}}$& 1.518\ \text{GeV}  & \cite{ParticleDataGroup:2024cfk}  \\
        $m_{\Lambda^{*}_{+}}$& 1.872\ \text{GeV}  &   \cite{ParticleDataGroup:2024cfk}\\
        $\lambda_{-}$& $(3.35\pm 0.15)\times 10^{-2}$\ \text{GeV$^3$}&\cite{Azizi:2024mmb} \\
        $m_{s}(\mu=1\ \text{GeV})$&0.126\ \text{GeV}  &\cite{ParticleDataGroup:2024cfk}   \\
        $f^{(1)}_{\Lambda_b}(\mu=1\ \text{GeV})$& $(0.030\pm 0.005)$\ \text{GeV$^3$}&\cite{Groote:1997yr} \\
        $f^{(2)}_{\Lambda_b}(\mu=1\ \text{GeV})$& $(0.030\pm 0.005)$\ \text{GeV$^3$} &\cite{Groote:1997yr} \\
       \hline
       \hline
       $\tau_{\Lambda_{b}}$&$1.471\times 10^{-12}\ \text{s}$&\cite{ParticleDataGroup:2024cfk}\\
       $m_{e}$&0.511 MeV&\cite{ParticleDataGroup:2024cfk}\\
       $m_{\mu}$&0.106 GeV &\cite{ParticleDataGroup:2024cfk}\\
       $m_{\tau}$&1.777 GeV&\cite{ParticleDataGroup:2024cfk}\\
       $m_B$&5.279 GeV&\cite{ParticleDataGroup:2024cfk}\\
       $m_K$&0.494 GeV&\cite{ParticleDataGroup:2024cfk}\\
       \hline
       \hline
     \end{tabular}    
 \end{table}
In addition, two auxiliary parameters are required for the sum rules of the form factors: the Borel parameter $M^2$ and the effective threshold parameter $s_0$. Ideally, the results of the form factors should be independent of these parameters. However, in practice, we can only determine specific regions where the dependence of the form factors on them is relatively weak. Moreover, the parameters $s_0$ and $M^2$ must be selected to suppress contributions from continuum states and higher-order twist $\Lambda_{b}$-LCDAs. To satisfy these conditions, the working regions for the Borel parameter $M^2$ and the effective threshold parameter $s_0$ are chosen as follows:
\begin{eqnarray}
\label{equ::M^2&S_0}
	&M^2=(3.5\pm 0.5)\ \text{GeV}^{2}\ ,\ \ \ \ \ 
	s_0=(3.6\pm 0.1)\ \text{GeV}^{2} \ ,
\end{eqnarray}
where the obtained regions of the Borel parameter $M^2$ and the effective threshold parameter $s_0$ are consistent with those used in \cite{Azizi:2024mmb}. To illustrate the numerical behavior of the LCSR predictions, we show the dependence of $f^{V}_{t}(0)$ on both the Borel parameter $M^2$ and the effective threshold parameter $s_0$ as an example in Fig.\ref{fig::dependence on M^2&s_0}. Similar dependencies are also observed for the other $\Lambda_{b}\rightarrow \Lambda(1520)$ transition form factors.
As shown in Fig.\ref{fig::dependence on M^2&s_0}, the results exhibit only a weak dependence on these two auxiliary parameters. This indicates that the conditions mentioned earlier have been effectively satisfied, confirming that the chosen working regions for both parameters are appropriate.

Using the values of all input parameters and the analytical expressions for the LCSR-predicted $\Lambda_{b}\rightarrow \Lambda(1520)$ form factors, the numerical results at $q^2=0$ are presented in Table \ref{Tab::results at q^2=0}, alongside results from other theoretical methods. The uncertainties in the form factors arise from variations in the Borel parameter $M^2$, the effective threshold parameter $s_0$, the decay constants of the $\Lambda(1520)$ and $\Lambda_{b}$ baryons $\lambda_{-},\ f^{(1,2)}_{\Lambda_b}$, as well as the parameter $\omega_{0}$ associated with the $\Lambda_{b}$-LCDAs.
\begin{table}
\caption{Theoretical numerical results of the $\Lambda_{b}\rightarrow \Lambda(1520)$ form factors at $q^2=0$ by using LCSR as well as other theoretical methods.}
\label{Tab::results at q^2=0}
\centering
\begin{tabular}{|c@{\hspace{0.5em}}|c@{\hspace{1em}} c c c@{\hspace{1em}}| }
\hline
\hline
    & This Work& LFQM\cite{Li:2022nim} & NRQM\cite{Mott:2015zma}& LQCD\cite{Meinel:2021mdj}\\
\hline
$f^{V}_{t}(0)$&$0.038\pm0.015$ &$0.051\pm 0.007$& 0.0029 &$-0.152\pm0.053$\\ 
$f^{V}_{0}(0)$&$0.038\pm0.015$&$0.051\pm 0.007$ & 0.0029& $0.071\pm0.008$\\
$f^{V}_{\perp}(0)$&$0.047\pm0.019$ &$0.067\pm 0.009$ & 0.0042& $0.109\pm0.015$\\
$f^{V}_{g}(0)$&$0.000\pm0.000$&$0.0123\pm0.0001$ &-0.0002 &$-0.039\pm0.014$\\
$g^{A}_{t}(0)$&$0.038\pm0.015$ &$0.044\pm0.005$ & 0.0031 &$0.071\pm0.006$\\
$g^{A}_{0}(0)$& $0.038\pm0.015$& $0.044\pm 0.005$& 0.0031 &$-0.128\pm0.047$\\
$g^{A}_{\perp}(0)$&$0.032\pm0.013$&$0.038\pm0.004$ & 0.0033 &$-0.126\pm0.047$\\
$g^{A}_{g}(0)$&$0.000\pm0.000$ &$0.020\pm0.004$ & 0.0004 &$0.009\pm0.083$\\
$f^{T}_{0}(0)$&$0.046\pm0.019$&$0.0294\pm 0.0007$ & 0.0038 &$0.099\pm0.015$\\
$f^{T}_{\perp}(0)$&$0.037\pm0.015$ &$0.046\pm0.006$ &0.0030 &$0.069\pm0.008$\\
$f^{T}_{g}(0)$&$0.000\pm0.000$ &$0.032\pm0.006$ & -0.0041 &$0.013\pm0.037$\\
$g^{T5}_{0}(0)$&$0.032\pm0.013$ &$0.040\pm0.005$ & 0.0032 &$-0.121\pm0.050$\\
$g^{T5}_{\perp}(0)$& $0.038\pm0.015$& $0.038\pm 0.004$& 0.0030 &$-0.133\pm0.047$\\
$g^{T5}_{g}(0)$& $0.000\pm0.000$&$-0.19\pm 0.05$ & 0.0072 &$1.167\pm0.688$\\
\hline
\hline
\end{tabular}
\end{table}
Before discussing the results, let us first briefly examine the implications of the heavy quark limit $m_b\rightarrow \infty$ and the large-recoil limit $q^2\rightarrow 0$, where the Soft-Collinear Effective Theory (SCET) is particularly applicable \cite{Charles:1998dr, Beneke:2000wa, Bauer:2000yr, Mannel:2011xg, Wang:2011uv}. In the framework of SCET, the hadronic matrix elements that define the $\Lambda_{b}\rightarrow \Lambda(1520)$ form factors in Eqs.(\ref{FF:vector})-(\ref{TFF:axial}) can be re-expressed in terms of a single remaining form factor, $\zeta(q^2)$, at leading order in $\alpha_{s}$ and $\Lambda_{QCD}/m_b$: 
\begin{align}
    \label{equ::SCET}
    \langle\Lambda^*(p)|\bar{s}\Gamma b|\Lambda_{b}(p+q)\rangle=\bar{u}^{\alpha}_{
    \Lambda^{*}}(p)v_{\alpha}[\zeta(q^2)]\Gamma u_{\Lambda_{b}}(p+q)\ ,
\end{align}
In the large-recoil limit $q^2\rightarrow 0$, we have:
\begin{align}
    \label{equ::large-recoil-limit}
    f^{V}_{t}(0)&=f^{V}_{0}(0)=f^{V}_{\perp}(0)=g^{A}_{t}(0)=g^{A}_{0}(0)=g^{A}_{\perp}(0) \nonumber\\
    &=f^{t}_{0}(0)=f^{T}_{\perp}(0)=g^{T5}_{0}(0)=g^{T5}_{\perp}(0)=\zeta(0)/m_{\Lambda_{b}}\ ,
\end{align}
whereas all $f(g)^{d}_{g}(q^2)$ form factors vanish also in the large-recoil limit.

We now present our comments on the numerical results for the form factors of the $\Lambda_{b}\rightarrow \Lambda(1520)$ transition, as predicted using the LCSR approach, which are summarized in Table \ref{Tab::results at q^2=0}.
\begin{itemize}
    \item At tree level, the LCSR approach yields zero results for all the $f(g)^{d}_{g}(q^2)$ form factors. This is because these form factors are only determined by the coefficient functions $\Pi^{d}_{1,2}(p^2,q^2)$, which correspond to the Lorentz structures $g_{\lambda\mu}$ and $g_{\lambda\mu} \not q$. However, these specific Lorentz structures are not present in the partonic representation. This outcome is consistent with the SCET prediction at leading order in both $\alpha_{s}$ and $\Lambda_{QCD}/m_b$, where the four form factors $f(g)^{d}_{g}(q^2)$ are non-zero only after including the next-to-leading order corrections in both  $\alpha_{s}$ and $\Lambda_{QCD}/m_b$.
    \item It is evident that, within the margin of error, the results for the remaining form factors at tree level in the LCSR approach are nearly identical. This finding is consistent with the predictions of SCET at leading-order of $\alpha_{s}$ and $\Lambda_{QCD}/m_b$, as given in Eq. (\ref{equ::large-recoil-limit}). This result is reasonable, as the LCSR approach is particularly well-suited for calculating heavy-to-light transition form factors in the large-recoil region. Consequently, one would expect the outcomes derived from the LCSR method to match the SCET predictions. Additionally, the endpoint relations at $q^2=0$, given in Eqs.(\ref{equ;;end relations1}) and (\ref{equ;;end relations2}), have been numerically verified with high precision using the LCSR approach.
    \item The uncertainties in the form factors are primarily attributed to the variation in the $\Lambda_b$-LCDAs parameter $\omega_0$, indicating that the numerical results for the form factors are strongly dependent on the $\Lambda_b$-LCDAs parameters. Therefore, achieving a high level of accuracy in the determination of the $\Lambda_b$-LCDAs parameters is crucial for making precise predictions of the form factors.
\end{itemize}
For comparison, we also present the theoretical predictions for the $\Lambda_{b}\rightarrow \Lambda(1520)$ transition form factors at $q^2=0$ from other approaches in Table \ref{Tab::results at q^2=0}. The central values of the form factors predicted by the LCSR approach at $q^2=0$ are approximately 75\% of those obtained from the LFQM \cite{Li:2022nim}. However, considering the uncertainties, the predictions from both approaches are consistent with each other. In contrast, the predictions from the NRQM \cite{Mott:2015zma} are about an order of magnitude smaller than both our results and those from the LFQM. The results from LQCD \cite{Meinel:2021mdj} are larger than those from all other approaches. These LQCD results at $q^2=0$ are inferred using the extrapolation formula from Ref. \cite{Meinel:2021mdj}, but it is important to note that the validity of the LQCD results is confined to the region $16$ GeV$^2 \leq q^2 \leq (m_{\Lambda_{b}}-m_{\Lambda^*_-})^2  $. Therefore, the reliability of the LQCD results at $q^2=0$ may be questionable.

\begin{figure}
\centering
    \subfigure{\includegraphics[width=0.31\linewidth]{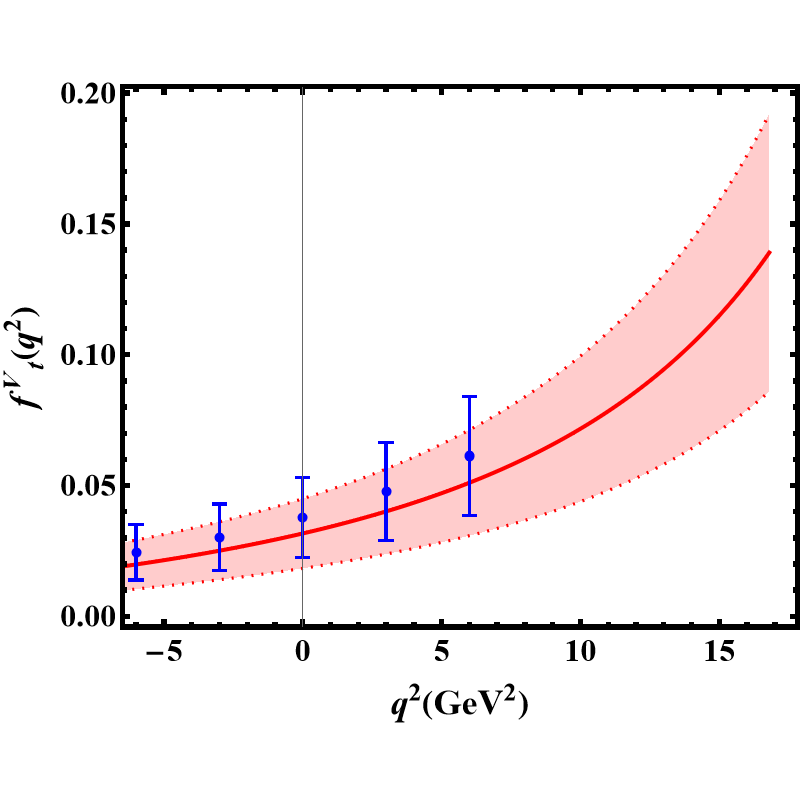}}
     \subfigure{\includegraphics[width=0.31\linewidth]{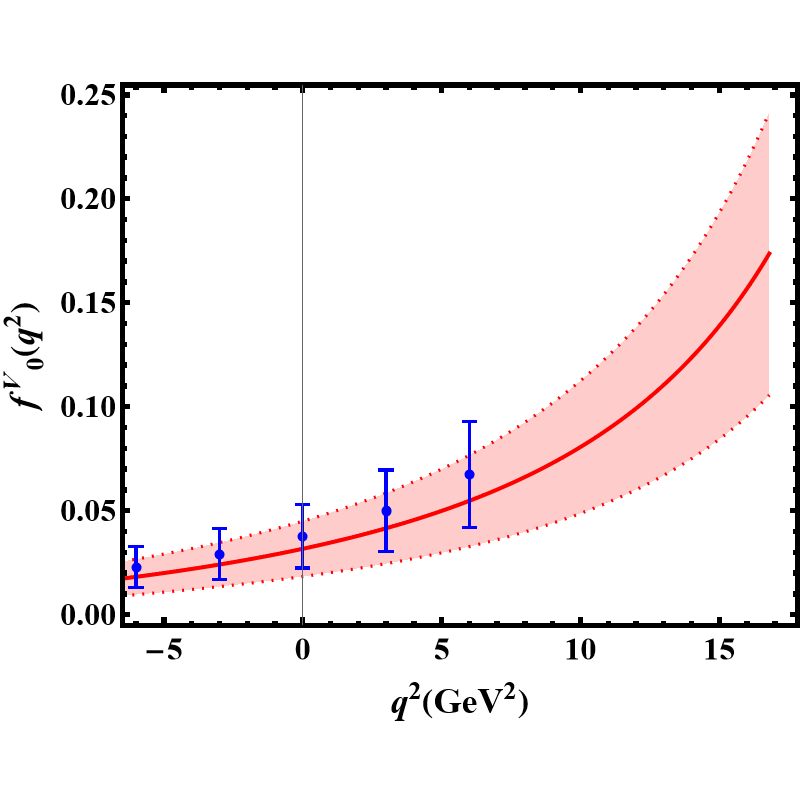}}
     \subfigure{\includegraphics[width=0.31\linewidth]{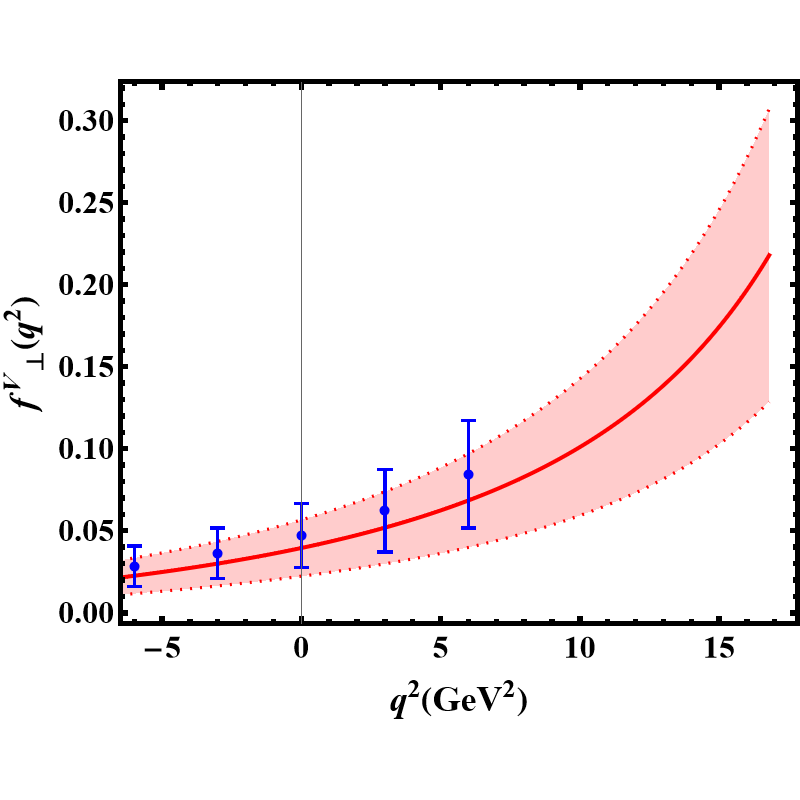}}
    \\
     \subfigure{\includegraphics[width=0.31\linewidth]{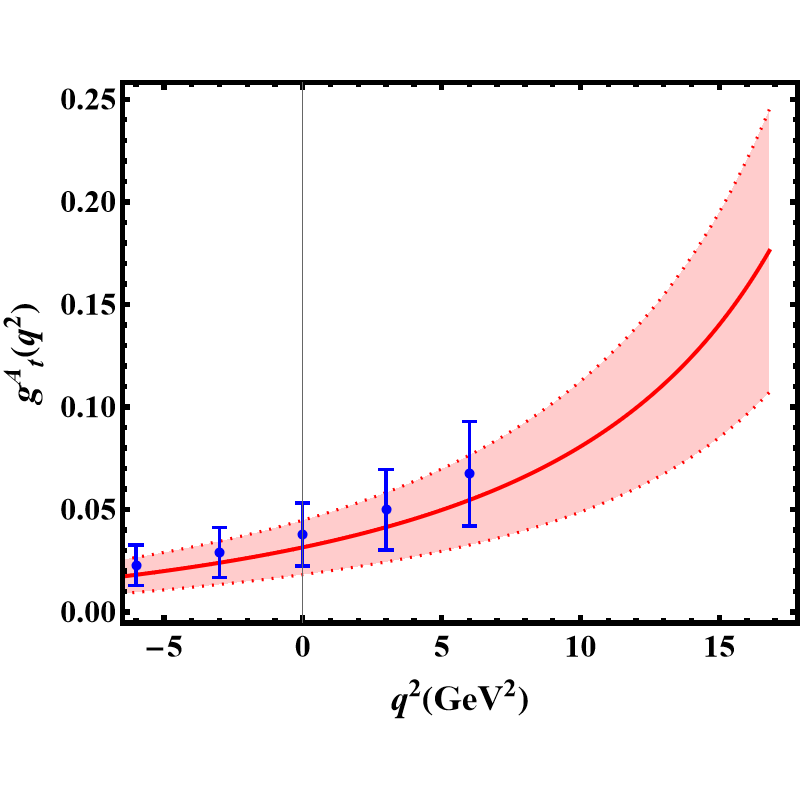}}
     \subfigure{\includegraphics[width=0.31\linewidth]{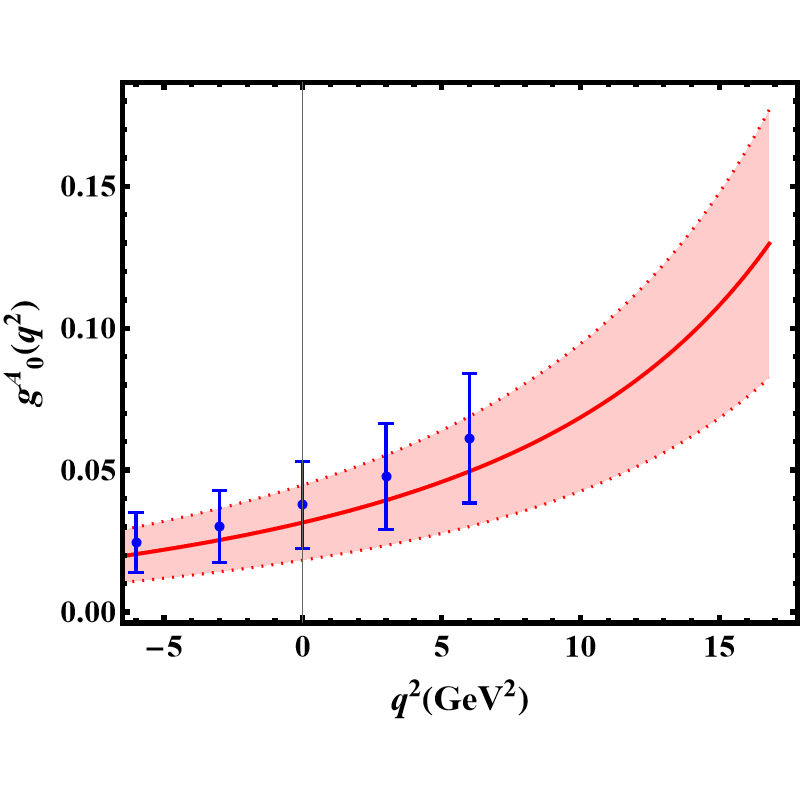}}
     \subfigure{\includegraphics[width=0.31\linewidth]{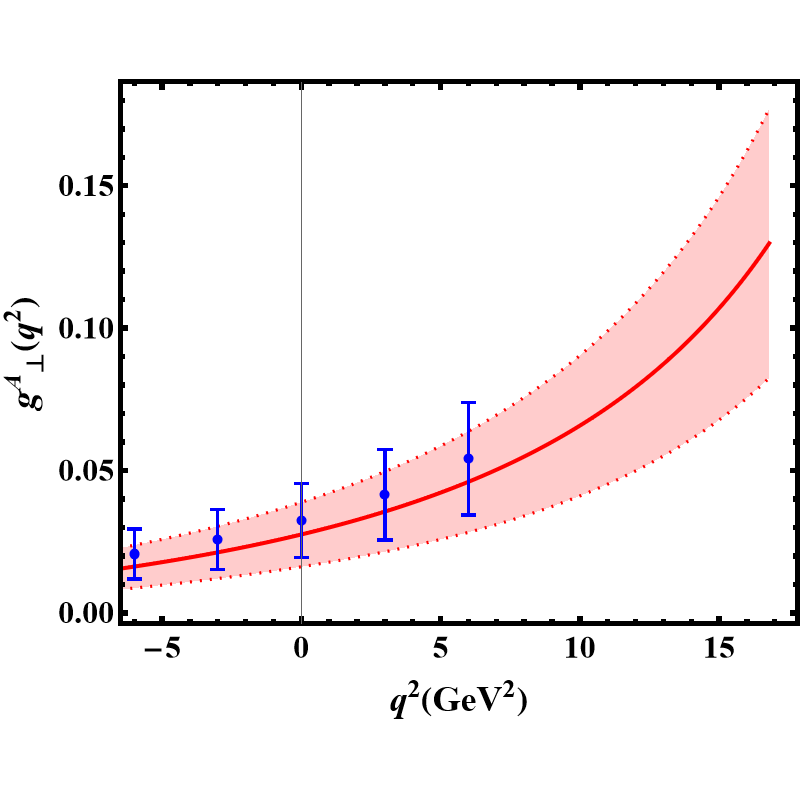}}
     \\
     \subfigure{\includegraphics[width=0.31\linewidth]{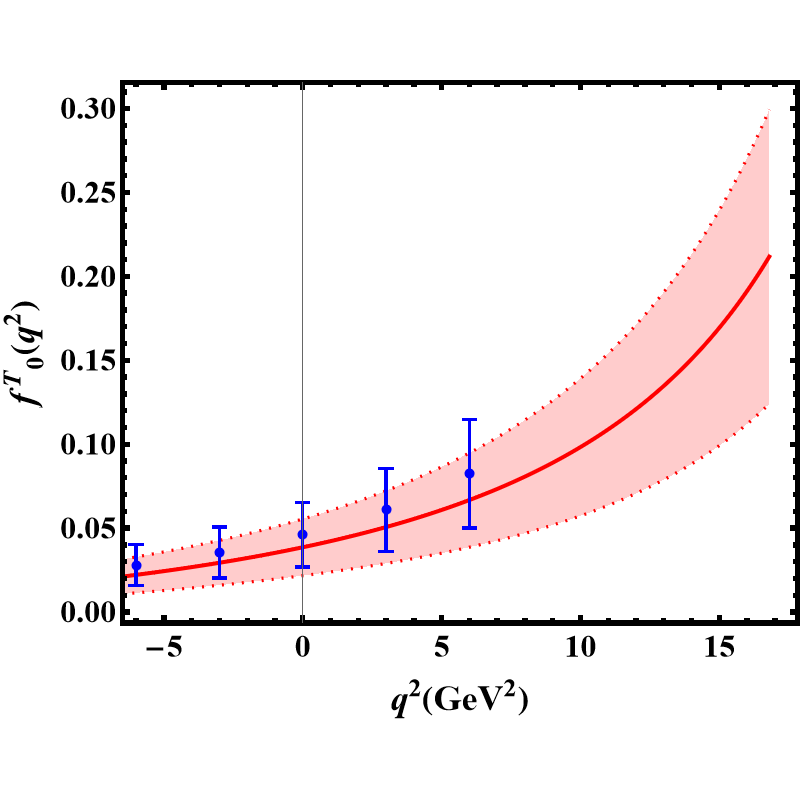}}
     \subfigure{\includegraphics[width=0.31\linewidth]{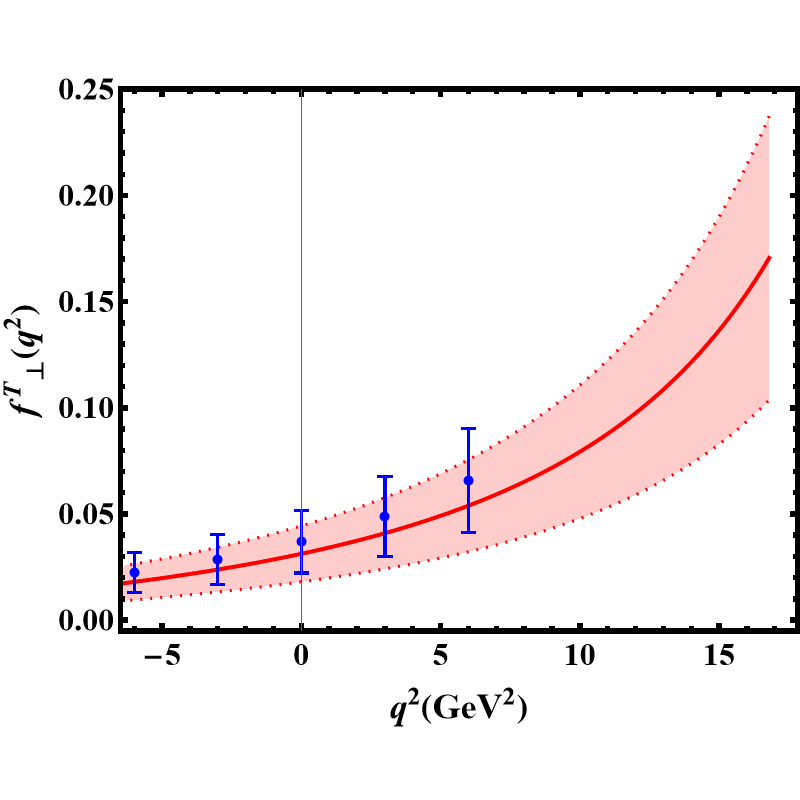}}
     \subfigure{\includegraphics[width=0.31\linewidth]{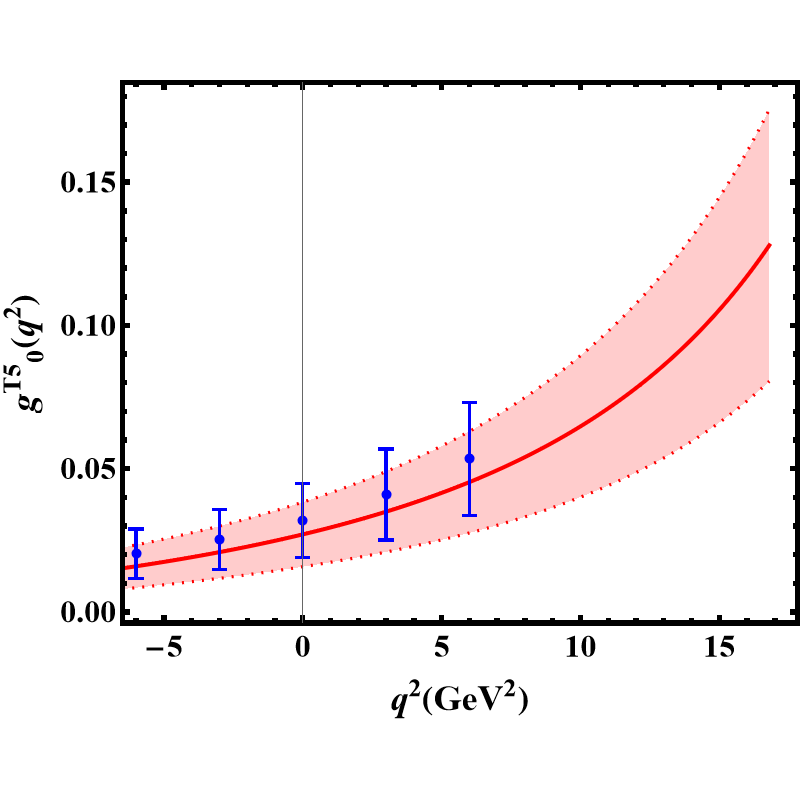}}
    \\
     \subfigure{\includegraphics[width=0.31\linewidth]{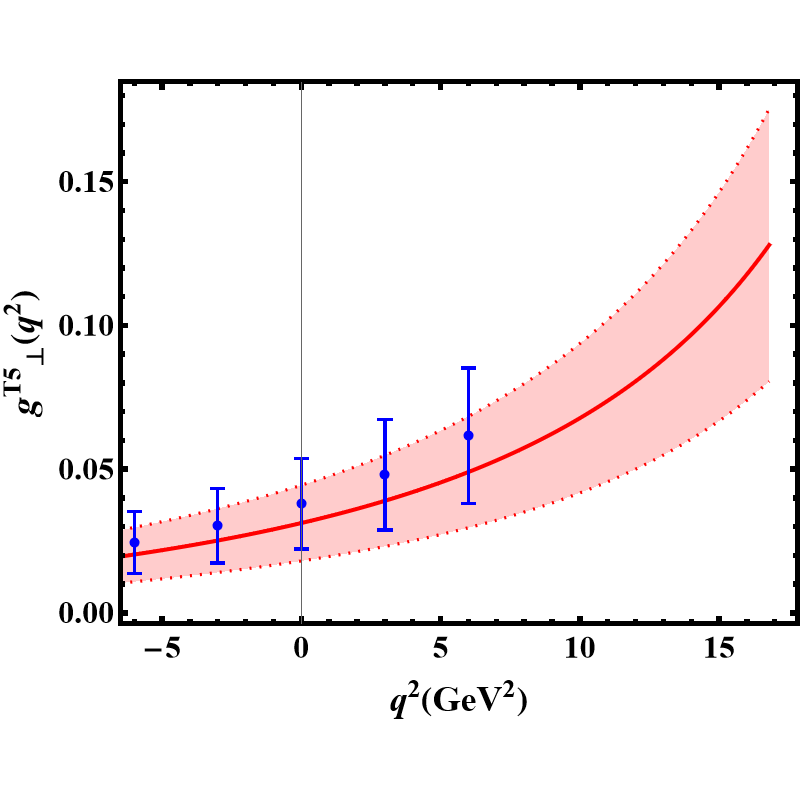}}
\caption{The $q^2$ dependence of the form factors of $\Lambda_{b}\rightarrow \Lambda(1520)$: where the red solid curves are the central values, the light red bands are the corresponding errors and the blue points are the samples used for the fitting at $q^2=\{-6.0,\ -3.0,\ 0.0,\ 3.0,\ 6.0\}$ Gev$^2$.}
\label{fig::dependence on a^2}
\end{figure}

The predictions from the $\Lambda_{b}$-baryon LCSR approach are reliable only within a limited range of momentum transfer, specifically for $q^2\leq 8\ \text{GeV}^2$, to ensure the validity of the light-cone operator-product expansion. Therefore, to extend the $\Lambda_{b}$-baryon LCSR calculation of the $\Lambda_{b}\rightarrow \Lambda(1520)$ transition form factors to larger values of $q^2$, we must perform an extrapolation using the $z$-series parametrization. This approach is based on the positive and analytic properties of the form factors \cite{Bourrely:2008za}. We apply the conformal transformation as outlined in Refs. \cite{Amhis:2022vcd, Blake:2022vfl} to achieve this extrapolation:
\begin{align}
    \label{equ::z-function}
    z(q^2,t_0)=\frac{\sqrt{t_+ - q^2}-\sqrt{t_+ - t_0}}{\sqrt{t_+ - q^2}+\sqrt{t_+ - t_0}}\ ,
\end{align}
with the threshold parameter $t_+\equiv (m_{B}+m_{K})^2$ for the $b\rightarrow s$ transitions. We choose $t_0=(m_{\Lambda_{b}}-m_{\Lambda^*_-})^2$ to make the entire physical momentum transfer $0<q^2<q^2_{max}\equiv (m_{\Lambda_{b}}-m_{\Lambda^*_-})^2$ map onto an interval of positive real $z$ axis. In order  to be the best parametrization of the form factors, we employ the following $z$-series expansion formula:
\begin{align}
    \label{equ::z-parameteration}
    f(q^2)=\frac{1}{1-q^2/(m^f_{\text{pole}})^2}\Big[a^{f}_{0}+a^{f}_{1}z(q^2,t_0)\Big]\ ,
\end{align} 
where the $m^f_{\text{pole}}$ for the different form factors are collected in Table. \ref{Tab::pole mass}.
 \begin{table}
  \caption{Table of $B_s$ meson pole mass appearing in the different form factors. The values are taken from \cite{ParticleDataGroup:2024cfk,Lang:2015hza}.}
  \label{Tab::pole mass}
  \centering
     \begin{tabular}{|c |c @{\hspace{1em}}c|}
     \hline
     \hline
     \text{From factors}&\text{Pole spin-parity $J^P$}&$m_{\text{pole}}$\text{(GeV)}\\
     \hline
  $f^{V}_{0},\ f^{V}_{\perp},\ f^{T}_{0},\ f^{T}_{\perp}$&$1^-$&5.416\\
  $f^{V}_{t}$&$0^+$&5.711\\
  $g^{A}_{0},\ g^{A}_{\perp},\ g^{T5}_{0},\ g^{T5}_{\perp}$&$1^+$&5.750\\
      $g^{A}_{t}$&$0^-$&5.367\\
      \hline
      \hline
     \end{tabular}    
 \end{table}
To determine the coefficients $a^{f}_{0,1}$ in the $z$-series expansion, as well as the correlation coefficients among them for each of the form factors involved, we perform a single $\chi^2$ fit using the $\Lambda_b$-baryon LCSR predictions for the $\Lambda_{b}\rightarrow \Lambda(1520)$ transition form factors at five distinct kinematic points for each form factor: $q^2=\{-6.0,\ -3.0,\ 0.0,\ 3.0,\ 6.0\}$ GeV$^2$. 
Before carrying out the fitting procedure, the endpoint relations given in Eqs. (\ref{equ;;end relations1}) and (\ref{equ;;end relations2}), which are independent of the specific method used to compute the form factors, allow us to derive precise relationships among the expansion coefficients. These relationships play a crucial role in reducing the number of fitting parameters required:
\begin{eqnarray}
    \label{equ::relation for a^f_i}
    a^{f^V_t}_0&&=a^{f^V_0}_0+a^{f^V_0}_1 z(0,t_0)-a^{f^V_t}_1z(0,t_0)\, ,\nonumber \\
     a^{g^A_t}_0&&=a^{g^A_0}_0+a^{g^A_0}_1 z(0,t_0)-a^{g^A_t}_1z(0,t_0)\, ,\nonumber \\
    a^{g^A_{\perp}}_0&&=a^{g^A_0}_0+a^{g^A_0}_1 z(q^2_{max},t_0)-a^{g^A_{\perp}}_1z(q^2_{max},t_0)\, ,\nonumber \\
    a^{f^T_{\perp}}_0&&=a^{g^{T5}_\perp}_0+a^{g^{T5}_0}_1 z(0,t_0)- a^{f^T_{\perp}}_1z(0,t_0)\, ,\nonumber \\
    a^{g^{T5}_0}_0&&=a^{g^{T5}_\perp}_0+a^{g^{T5}_0}_1 z(q^2_{max},t_0)- a^{g^{T5}_0}_1z(q^2_{max},t_0)\, .
\end{eqnarray} 
Since all four $f(g)^d_g(q^2)$ form factors are zero at tree level in the LCSR approach, we are left with five relations between the expansion coefficients $a^{f}_{0,1}$. To perform the fitting, we generate N = 500 ensembles of the input parameters, which include the Borel parameter $M^2$, the effective threshold parameter $s_0$, the decay constants $\lambda_-$ and $f^{(1,2)}_{\Lambda_b}$, as well as the parameter $\omega_0$ of the $\Lambda_b$-LCDAs. These ensembles are generated using a multivariate joint distribution \cite{SentitemsuImsong:2014plu, Huang:2022lfr}. From these N = 500  parameter sets, we obtain the samples necessary for the fitting process. The results of the fitting are presented in Table \ref{Table:a^f_0}
\begin{table}
    \caption{The fitting results of the $z$-series expansion coefficients,$a^f_0$ and $a^f_1$. The unlisted coefficients should be determined using Eq.(\ref{equ::relation for a^f_i}). The correlation coefficient matrix between the $z$-series expansion coefficients $a^f_{1,2}$ is given in Appendix.\ref{sec:Appendix-D}}
    \label{Table:a^f_0}
    \centering
    \begin{tabular}{|@{\hspace{0.7em}}c@{\hspace{0.7em}}|@{\hspace{0.7em}}c@{\hspace{1.5em}}c@{\hspace{0.7em}}|}
    \hline
    \hline
        From factor& $a^f_0$ &  $a^f_1$\\
    \hline
        $f^V_t$&  &$-0.206\pm 0.082$   \\
        $f^V_0$&$0.074\pm0.029$&$-0.245\pm0.098$   \\
        $f^V_\perp$&$0.093\pm0.038$&$-0.309\pm0.129$   \\
        $g^A_t$&  &$-0.241\pm0.097$   \\
        $g^A_0$& $0.064\pm0.023$ & $-0.186\pm0.064$  \\
        $g^A_\perp$&  & $-0.209\pm0.073$  \\
        $f^{T}_0$& $0.091\pm0.037$ &$-0.299\pm0.127$   \\
        $f^{T}_\perp$&  &$-0.240\pm0.095$   \\
        $g^{T5}_0$&  & $-0.207\pm0.074$  \\
        $g^{T5}_\perp$&$0.063\pm0.023$  & $-0.183\pm0.065$  \\
        \hline
        \hline
    \end{tabular}
\end{table}
, and the correlation coefficient matrix between these coefficients is given in Appendix.\ref{sec:Appendix-D}. The $q^2$ dependence of the form factors is shown in Fig.\ref{fig::dependence on a^2}.
The relative errors of the form factors are predominantly around 50 \%, primarily due to their strong dependence on the uncertainty of the $\Lambda_b$-LCDAs parameter $\omega_0$.

\subsection{Phenomenology}
We can now use the obtained $\Lambda_{b}\rightarrow \Lambda(1520)$ transition form factors to perform a phenomenological analysis. The effective Hamiltonian governing the $\Lambda_{b}\rightarrow \Lambda(1520)l^{+}l^{-}$ decay processes is given in Eq. (\ref{equ::Hamiltonian}). This Hamiltonian allows us to derive several physical observables associated with the $\Lambda_{b}\rightarrow \Lambda(1520)l^{+}l^{-}$ decay channels. Following Refs. \cite{Descotes-Genon:2019dbw, Das:2020cpv}, we can obtain the full expression for the 4-fold differential decay distribution for the process $\Lambda_{b}\rightarrow \Lambda(1520)l^{+}l^{-}$. 
It is important to note that in Ref. \cite{Descotes-Genon:2019dbw}, the lepton masses were neglected, while in Ref. \cite{Das:2020cpv}, they were included. In our calculation, we take into account the effect of the lepton mass. The differential angular distribution for the 4-fold decay $\Lambda_{b}\rightarrow \Lambda(1520)l^{+}l^{-}$ is given by the following formula:
\begin{align}
    \label{equ::differential angular distribution}
    \frac{d4\Gamma}{dq^2d \cos{\theta_{\Lambda^*}}d \cos{\theta_l}d\phi}&=\frac{3}{8\pi}\sum_{i}L_i(q^2)f_i(q^2,\cos{\theta_{\Lambda^*}},\cos{\theta_l},\phi) \nonumber \\
    &=\frac{3}{8\pi}\Big[\big(L_{1c}\cos{\theta_l}+L_{1cc}\cos^2{\theta_l}+L_{1ss}\sin^2{\theta_l}\big)\cos^2{\theta_\Lambda^*}\nonumber\\
    &\ \ \ \ +\big(L_{2c}\cos{\theta_l}+L_{2cc}\cos^2{\theta_l}+L_{2ss}\sin^2{\theta_l}\big)\sin^2{\theta_\Lambda^*}\nonumber\\
     &\ \ \ \ +\big(L_{3ss}\sin^2{\theta_l}\cos^2{\phi}+L_{4ss}\sin^2{\theta_l}\sin{\phi}\cos{\phi}\big)\sin^2{\theta_\Lambda^*}\nonumber\\
     &\ \ \ \ +\big(L_{5s}\sin{\theta_l}+L_{5sc}\sin{\theta_l}\cos{\theta_l}\big)\sin{\theta_\Lambda^*}\cos{\theta_\Lambda^*}\cos{\phi}\nonumber\\
     &\ \ \ \ +\big(L_{6s}\sin{\theta_l}+L_{6sc}\sin{\theta_l}\cos{\theta_l}\big)\sin{\theta_\Lambda^*}\cos{\theta_\Lambda^*}\sin{\phi}\Big]\ ,
\end{align}
where the angular coefficients $L_i$ are functions of the momentum transfer $q^2$ alone, and their exact expressions in terms of the form factors are provided in Appendix-G of Ref. \cite{Das:2020cpv}. In our calculation, we do not include the branching ratio $\mathcal{B}_{\Lambda^*_-}\equiv \mathcal{B}(\Lambda^*_-\rightarrow N\bar{K})$ in the angular coefficients $L_i$. As a result, the physical observables defined here are directly related to the $\Lambda_{b}\rightarrow \Lambda(1520)l^{+}l^{-}$ decay processes.

Integrating over the  angles $\theta_l,\ \theta_\Lambda^*,\ \phi$ in Eq.(\ref{equ::differential angular distribution}), the relevant physical observables  can be defined with the following formula:
\begin{itemize}
    \item The differential branching fraction $d\mathcal{B}/dq^2$:
    \begin{align}
        \label{equ::dBF}
        \frac{d\mathcal{B}}{dq^2}&=\tau_{\Lambda_b}\times\int^1_{-1}d\cos{\theta_l}\int^1_{-1}d\cos{\theta_{\Lambda^*}}\int^{2\pi}_0 d\phi\frac{d4\Gamma}{dq^2d \cos{\theta_{\Lambda^*}}d \cos{\theta_l}d\phi} \nonumber \\
&=\tau_{\Lambda_b}\times\frac{L_{1cc}+2L_{1ss}+2L_{2cc}+4L_{2ss}+2L_{3ss}}{3}\ ,
    \end{align}
    \item The lepton-side forward-backward asymmetry $A_{FB}$:
    \begin{align}
        \label{equ::AlFB}
        A_{FB}&=\frac{(\int^0_{-1}-\int^1_{0})d\cos{\theta_l}\int^1_{-1}d\cos{\theta_{\Lambda^*}}\int^{2\pi}_0 d\phi\frac{d4\Gamma}{dq^2d \cos{\theta_{\Lambda^*}}d \cos{\theta_l}d\phi}}{\int^1_{-1}d\cos{\theta_l}\int^1_{-1}d\cos{\theta_{\Lambda^*}}\int^{2\pi}_0 d\phi\frac{d4\Gamma}{dq^2d \cos{\theta_{\Lambda^*}}d \cos{\theta_l}d\phi}} \nonumber \\
        &=\frac{3}{2}\frac{L_{1c}+2L_{2c}}{L_{1cc}+2L_{1ss}+2L_{2cc}+4L_{2ss}+2L_{3ss}}\ ,
    \end{align}
    \item The longitudinal polarization fractions of the di-lepton system $F_{L}$:
    \begin{align}
        \label{equ::FT}
        F_{L}=1-\frac{2(L_{1cc}+2L_{2cc})}{L_{1cc}+2L_{1ss}+2L_{2cc}+4L_{2ss}+2L_{3ss}}\ ,
    \end{align}
    \item the CP-averaged normalized angular observable$ S_{1cc}(q^2)$:
    \begin{align}
        \label{equ::S_1cc}
        S_{1cc}(q^2)=\frac{L_{1cc}+\bar{L}_{1cc}}{d(\Gamma+\bar{\Gamma})/dq^2}\ .
    \end{align}
    where the bar denotes CP-conjugated quantities and $\bar{L}_{i}(q^2)$ can be obtained by doing the full conjugation for all weak phases in $L_{i}(q^2)$.
    \end{itemize}
The physical observables defined above are the most promising candidates for measurement at the LHCb, as discussed in Ref. \cite{Amhis:2020phx}.

Using the effective Wilson coefficients $C^{eff}_{7}(m_b),\ C^{eff}_{9}(m_b,q^{2}),\ C_{10}(m_b)$ provided in Refs. \cite{Yan:2000dc, Azizi:2012vy, Li:2004vh, Buras:1994dj, Ali:1994bf} at the scale $\mu=m_b$, we can predict the physical observables discussed above. However, it is important to note that we only consider the factorizable quark-loop contributions in the effective Wilson coefficients $C^{eff}_{7}$ and $ C^{eff}_{9}(q^{2})$. Specifically, we include only the charm-loop contribution derived from perturbation theory (with $m_c=1.4$ ) GeV) and exclude the non-factorizable long-distance effects from the $c\bar{c}$ resonant region. 
In the experimental measurement \cite{LHCb:2023ptw}, contributions from the $c\bar{c}$ resonant region, such as the $J/ \Psi$ and $\psi(2s)$ charmonium resonances, are specifically excluded from the spectrum. This allows our predictions to be compared more directly with experimental results, with minimal influence from the complexities associated with non-factorizable long-distance effects.

\begin{figure}
\centering
    \subfigure{\includegraphics[width=0.31\linewidth]{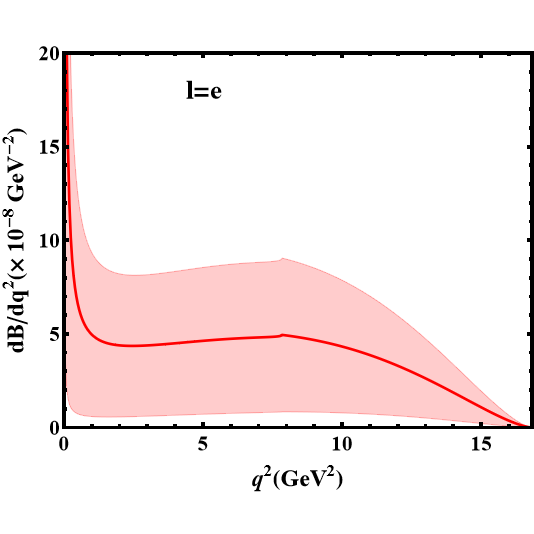}}
     \subfigure{\includegraphics[width=0.31\linewidth]{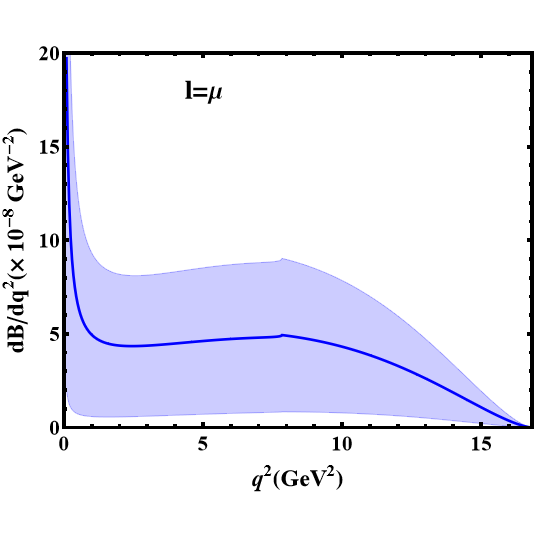}}
     \subfigure{\includegraphics[width=0.31\linewidth]{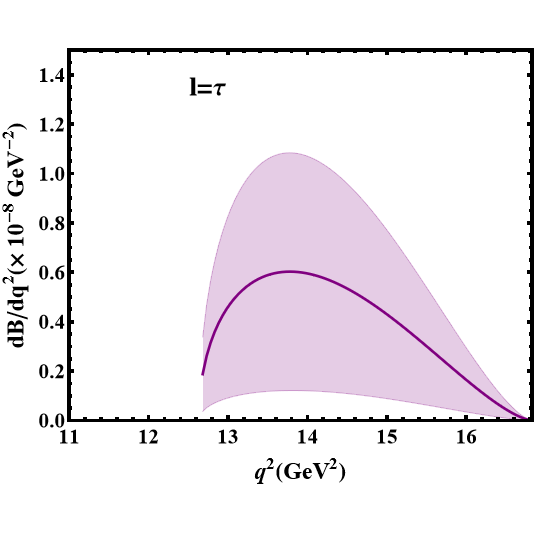}}
    \\
\caption{The $q^2$ dependence of the differential branching fraction $d\mathcal{B}/dq^2$ of $\Lambda_{b}\rightarrow \Lambda(1520)l^+ l^-(l=e(\text{left}),\ \mu(\text{center}), \ \tau(\text{right}))$: where the red, the blue and the purple solid curves are the central values of our prediction for the $e,\ \mu\ \text{and}\  \tau$ channels, respectively, and the red, blue and purple light bands are the corresponding errors.}
\label{fig::dependence on dBr}
\end{figure}

\begin{figure}
\centering
    \subfigure{\includegraphics[width=0.31\linewidth]{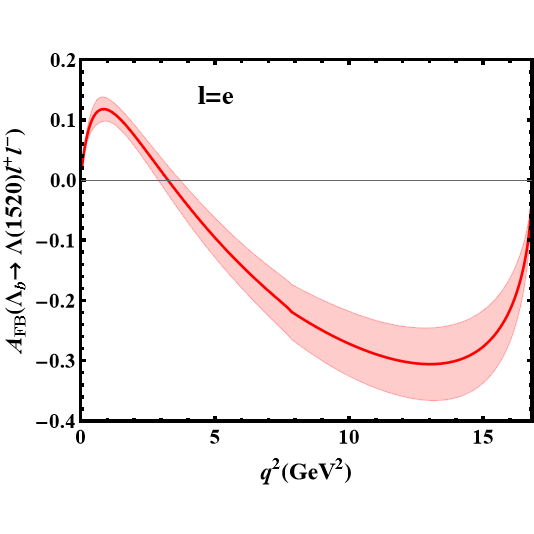}}
     \subfigure{\includegraphics[width=0.31\linewidth]{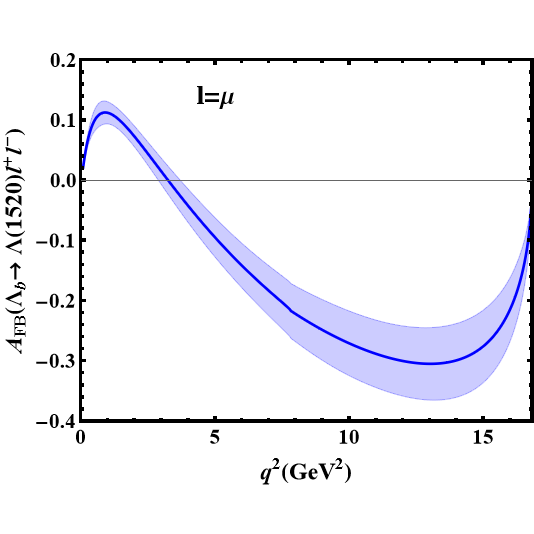}}
     \subfigure{\includegraphics[width=0.31\linewidth]{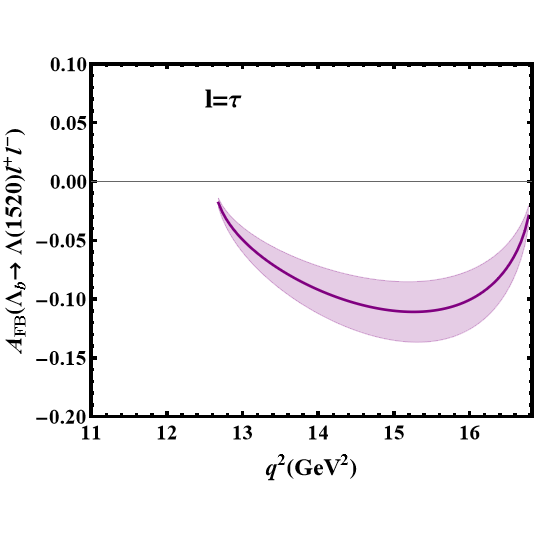}}
\caption{The $q^2$ dependence of the lepton-side forward-backward asymmetry $A_{FB}$ of $\Lambda_{b}\rightarrow \Lambda(1520)l^+ l^-(l=e(\text{left}),\mu(\text{center}), \tau(\text{right}))$: where the red, the blue and the purple solid curves are the central values of our prediction for the $e,\ \mu\ \text{and}\  \tau$ channels, respectively, and the red, blue and purple light bands are the corresponding errors.}
\label{fig::dependence on dAFB}
\end{figure}

\begin{figure}
\centering
    \subfigure{\includegraphics[width=0.31\linewidth]{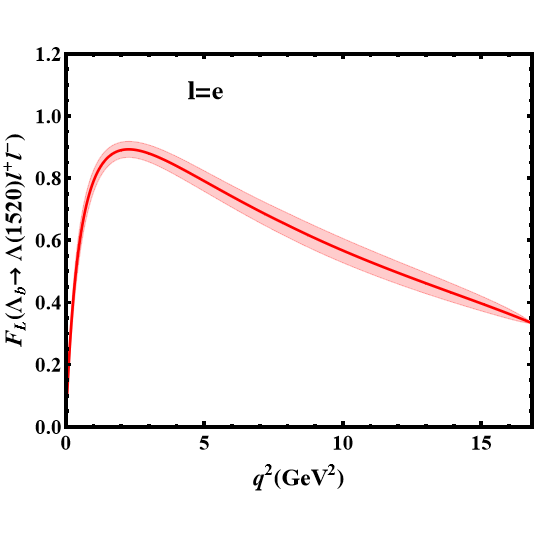}}
     \subfigure{\includegraphics[width=0.31\linewidth]{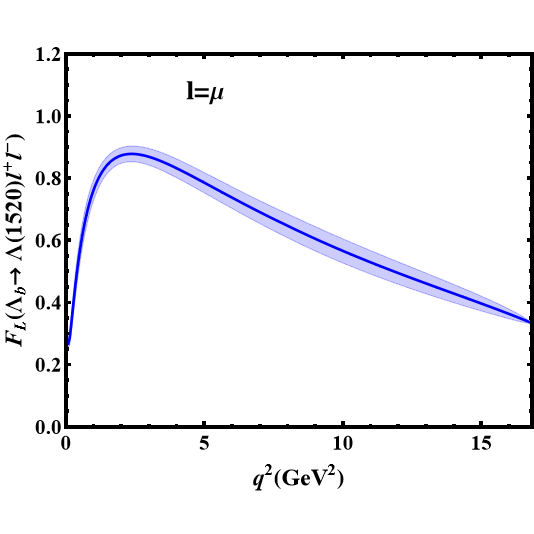}}
     \subfigure{\includegraphics[width=0.31\linewidth]{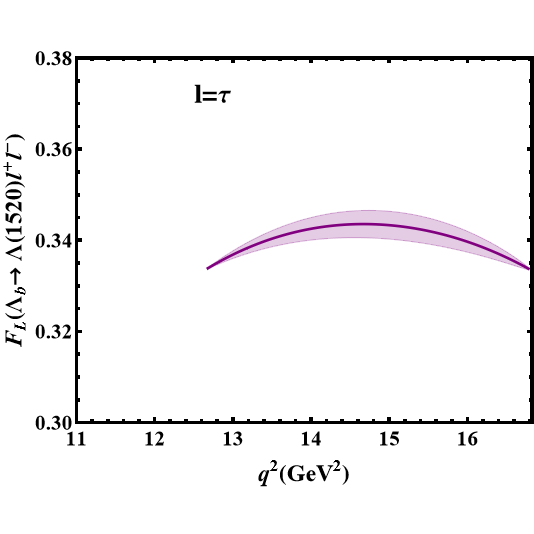}}
\caption{The $q^2$ dependence of the longitudinal polarization fractions $F_{L}$ of $\Lambda_{b}\rightarrow \Lambda(1520)l^+ l^-(l=e(\text{left}),\ \mu(\text{center}),\ \tau(\text{right}))$: where the red, the blue and the purple solid curves are the central values of our prediction for the $e,\ \mu\ \text{and}\  \tau$ channels, respectively, and the red, blue and purple light bands are the corresponding errors.}
\label{fig::dependence on dFL}
\end{figure}

\begin{figure}
\centering
    \subfigure{\includegraphics[width=0.31\linewidth]{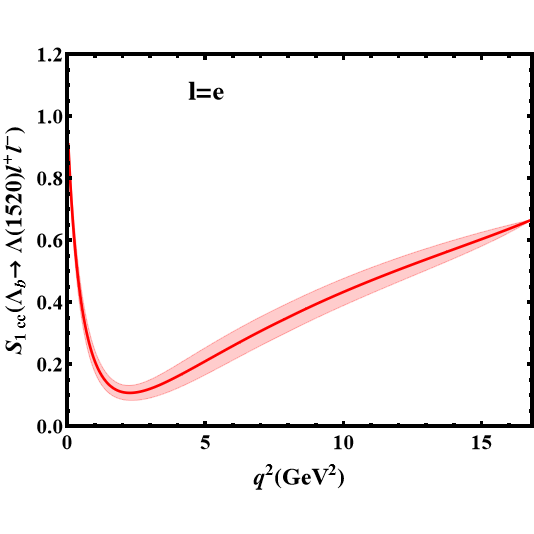}}
     \subfigure{\includegraphics[width=0.31\linewidth]{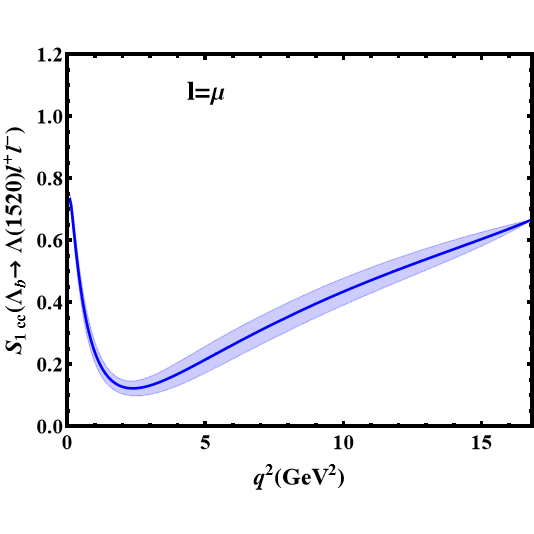}}
     \subfigure{\includegraphics[width=0.31\linewidth]{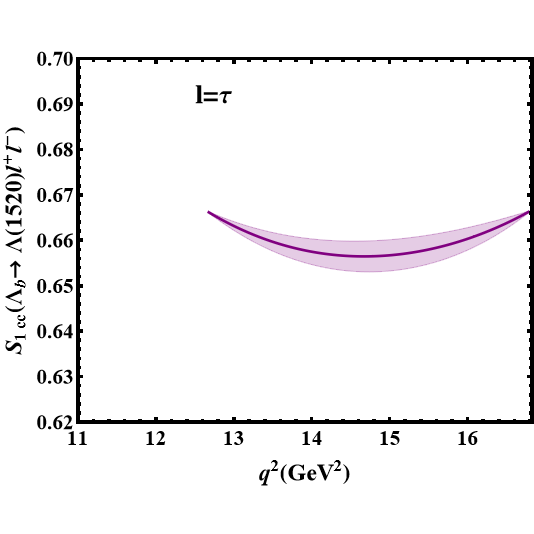}}
\caption{The $q^2$ dependence of the CP-averaged normalized angular obesrvable $S_{1cc}$ of $\Lambda_{b}\rightarrow \Lambda(1520)l^+ l^-(l=e(\text{left}),\ \mu(\text{center}),\ \tau(\text{right}))$: where the red, the blue and the purple solid curves are the central values of our prediction for the $e,\ \mu\ \text{and}\  \tau$ channels, respectively, and the red, blue and purple light bands are the corresponding errors.}
\label{fig::dependence on ds1cc}
\end{figure}

Now, we present our predictions for the $q^2$-dependence of the physical observables discussed above, using the form factors predicted by the $\Lambda_b$-LCSR approach, as derived in the previous subsection. The $q^2$-dependence of the differential branching fraction $d\mathcal{B}/dq^2$, the lepton-side forward-backward asymmetry $A_{FB}$, the longitudinal polarization fraction $F_{L}$, and the CP-averaged normalized angular observable $S_{1cc}$ are shown in Figs. \ref{fig::dependence on dBr}, \ref{fig::dependence on dAFB}, \ref{fig::dependence on dFL}, and \ref{fig::dependence on ds1cc}, respectively.

\begin{figure}
\centering
    \subfigure{\includegraphics[width=0.53\linewidth]{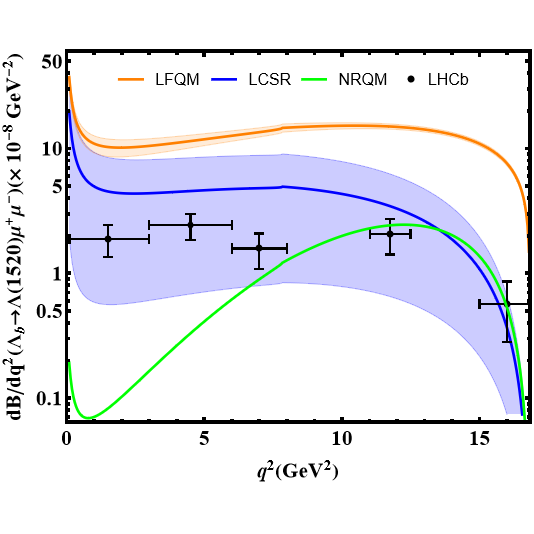}}
    \\
\caption{For comparison, we give the $q^2$ dependence of the differential branching $d\mathcal{B}/dq^2$ for the $\Lambda_{b}\rightarrow \Lambda(1520)\mu^+\mu^-$ from other theoretical and experimental predictions: the LFQM \cite{Li:2022nim}(the yellow curve and band), the LCSR(this work)(the blue curve and band), the NRQM \cite{Mott:2015zma}(the green curve) and the LHCb \cite{LHCb:2023ptw}(the black point).}
\label{fig::dependence on dBr-mu-compare}
\end{figure}
For comparison, we also present the results of the differential branching fraction $d\mathcal{B}/dq^2$ of the $\Lambda_b\rightarrow \Lambda(1520)\mu^+\mu^-$ process from the experimental measurement and other theoretical approaches in Fig.\ref {fig::dependence on dBr-mu-compare}. 
The LHCb collaboration has measured the differential branching fraction $d\mathcal{B}/dq^2$ for the $\Lambda_b\rightarrow \Lambda(1520)\mu^+\mu^-$ process for the first time, as reported in Ref. \cite{LHCb:2023ptw}. Using the form factors for $\Lambda_b\rightarrow \Lambda(1520)$ calculated within the LFQM \cite{Li:2022nim} and the NRQM \cite{Mott:2015zma}, we can also predict the differential branching fraction $d\mathcal{B}/dq^2$ of $\mu$ channel. 
It should be noted that the form factors for $\Lambda_b\rightarrow \Lambda(1520)$ calculated using NRQM \cite{Mott:2015zma} are presented without any associated uncertainties, and only the central values are available. Therefore, we will only show the central values of the differential branching fraction $d\mathcal{B}/dq^2$ obtained from NRQM \cite{Mott:2015zma}. 
Additionally, we do not include the results from LQCD \cite{Meinel:2021mdj} in our predictions, as the form factors calculated by LQCD are valid only within the region $16$ GeV$^2\leq q^2 \leq (m_{\Lambda_{b}}-m_{\Lambda^*_-})^2  $.

We now provide a detailed discussion and analysis of the results for these physical observables, which are presented in Figs. \ref{fig::dependence on dBr},  \ref{fig::dependence on dAFB}, \ref{fig::dependence on dFL} , \ref{fig::dependence on ds1cc} and \ref{fig::dependence on dBr-mu-compare}
\begin{itemize}
\item In Figs. \ref{fig::dependence on dBr}, \ref{fig::dependence on dAFB}, \ref{fig::dependence on dFL}, \ref{fig::dependence on ds1cc}, and \ref{fig::dependence on dBr-mu-compare}, mild kinks are observed at $q^2=4m_c^2$, which are more obvious in the plot of the differential branching fraction $d\mathcal{B}/dq^2$ and less pronounced in the figures for the lepton-side forward-backward asymmetry $A_{FB}$, the longitudinal polarization fraction $F_{L}$, and the CP-averaged normalized angular observable $S_{1cc}$. These kinks correspond to the $c\bar{c}$ threshold at $q^2=4m_c^2$, arising from the perturbative charm-loop contribution to the effective Wilson coefficient $C^{eff}_9(q^2)$, which introduces an imaginary part for $q^2\geq 4m_c^2$.
\item The relative errors of the differential branching fraction $d\mathcal{B}/dq^2$ predicted by the the form factors of LCSR at large-recoil region are almost 80\%. This is because of the large relative errors of the form factors predicted by LCSR, which are connected with the high sensitivity to the error of the $\Lambda_b$-LCDAs parameter $\omega_0$. Therefore, the high-precision $\Lambda_b$-LCDAs parameters are urgently needed for the high-precision prediction of the $\Lambda_b\rightarrow \Lambda(1520)$ transition from factors. As for the rest of physical observables, because of the definition form of division in Eqs.(\ref{equ::AlFB}) - (\ref{equ::S_1cc}), the relative errors of these physical observables are smaller. Additionally, the strong correlation between the $z$-series expansion coefficients $a^f_{0,1}$ have reduced part of the errors for these physical observables.
\item We have numerically found that the dominant contributions to the differential branching fraction in large-recoil region are mainly from the form factors $f^V_0(q^2)$ and $g^A_0(q^2)$, and sub-dominant contributions are from the the form factors $f^{V,T}_{\perp}(q^2)$ and $g^{A,T5}_{\perp}(q^2)$. We also checked that the contribution from the form factors $f^{V,T}_g(q^2)$ and  $g^{A,T5}_g(q^2)$ have very few influence on the differential branching fraction in large-recoil region by using the results of the from factors from the LFQM \cite{Li:2022nim} and NRQM \cite{Mott:2015zma}.
\item In Fig.\ref {fig::dependence on dBr-mu-compare} for the $\mu$ channel, the result of the differential branching fraction based on the from factors form the NRQM \cite{Mott:2015zma} is approximately one order of magnitude smaller than the experimental result of LHCb \cite{LHCb:2023ptw} in large-recoil region, and the result in the low-recoil is consistent with that of the LHCb. The result based on the form factors from the LFQM \cite{Li:2022nim} is  about five time larger than the experimental result of LHCb \cite{LHCb:2023ptw}. The central value of the differential branching fraction based on the form factors from this work is just about one time larger than that of LHCb \cite{LHCb:2023ptw} in large-recoil region, and consistent with the experimental result very well in low-recoil region. If taking the errors into account, our work stands as the most consistent with the current experimental results, outperforming other theoretical works in this regard. This underscores the reliability of the LCSR as a robust method for calculating heavy-to-light transition form factors in the large-recoil region. Once the next leading order corrections in both $\alpha_{s}$ and $\Lambda_{QCD}/m_b$  are considered, along with the inclusion of high-precision LCDAs parameters, the prediction based on the LCSR will undergo significant improvement.
\item As for the lepton-side forward-backward asymmetry $A_{FB}$, the longitudinal polarization fractions $F_{L}$ and the CP-averaged normalized angular observable $S_{1cc}$, the behaviors of the $q^2$ dependence for these physical observables are similar with that from the other theoretical works in most of the $q^2$ region, such as, the NRQM \cite{Mott:2015zma}, the LQCD \cite{Meinel:2020owd,Meinel:2021mdj} and the dispersive analysis \cite{Amhis:2022vcd}. However, compared with the results of the LQCD \cite{Meinel:2020owd,Meinel:2021mdj}, the NRQM \cite{Mott:2015zma}and the dispersive analysis \cite{Amhis:2022vcd}, our result for the the lepton-side forward-backward asymmetry $A_{FB}$ show only one zero-crossing in the large-recoil region, which is due to a combination of the effective Wilson coefficients $C^{eff}_7$ and $C^{eff}_9(q^2)$ that vanishes at some $q^2$ point. And there is no second zero-crossing in the low-recoil region. This is  because of that our result of the form factor $f^{V}_{\perp}(q^2)$ is larger than zero and does not cross the zero axis in the entire physical $q^2$ region. Therefore, we hope that the future experimental measurement of the lepton-side forward-backward asymmetry $A_{FB}$ in low-recoil region can tell us which kind of the behavior is correct.
\end{itemize}
\section{Summary}
\label{Summary}
In this work, we have calculated the form factors of $\Lambda_b \rightarrow \Lambda(1520)$ transition within the framework of LCSR with the LCDAs of $\Lambda_b$-baryon, then predicted some physical observables with the obtained form factors, such as the differential branching fraction $d\mathcal{B}/dq^2$, the lepton-side forward-backward asymmetry $A_{FB}$, the longitudinal polarization fractions $F_{L}$ and the CP-averaged normalized angular observable $S_{1cc}$ of the baryonic decay $\Lambda_b\rightarrow \Lambda(1520)l^+l^-$. By considering the special Lorentz structures and following the standard procedure of calculating the heavy-to-light form factors by utilizing the LCSR approach, we obtained the sum rules of the $\Lambda_b \rightarrow \Lambda(1520)$ transition form factors. In order to exclude the contamination of the positive-parity baryon $\Lambda(1890)$, we have included both the negative-parity baryon $\Lambda(1520)$ and the positive-parity baryon $\Lambda(1890)$ in the hadronic representation of the correlation function, therefore the $\Lambda_b \rightarrow \Lambda(1520)$ transition form factors could be evaluated exactly by matching the coefficients of the eight Lorentz structures in both the hadronic and partonic representations and solving linear equations. 

We have found that the results of the $\Lambda_b \rightarrow \Lambda(1520)$ transition form factors, predicted by LCSR in the large-recoil region, fully  comply with the relations predicted by SCET at leading order in $\alpha_{s}$ and $\Lambda_{QCD}/m_b$, and all four  $f(g)^{d}_{g}(q^2)$ form factors are zero. This demonstrates the reliability of the LCSR for calculating the heavy-to-light form factors in the large-recoil region. We have also compared our results of the $\Lambda_b \rightarrow \Lambda(1520)$ transition form factors at $q^2=0$ with that of other theoretical approaches. The results of the NRQM \cite{Mott:2015zma} are an order of magnitude smaller than the results of our work, and our results are about 75\% of that based on the LFQM \cite{Li:2022nim}. The results predicted by the LQCD \cite{Meinel:2021mdj} are an order of magnitude larger than that of all of the rest works. However, the reliability of the results in the large-recoil region based on the LQCD \cite{Meinel:2021mdj} should be questionable. It also should be mentioned that the errors of our results are sizable, mainly due to the high sensitivity to the error of the $\Lambda_b$-LCDAs parameter $\omega_0$. Therefore, more attention should be paid to constraining the parameters of the $\Lambda_b$-LCDAs. In order to extrapolate the LCSR-predicted results to the whole physical region, we have employed the $z$-series expansion formula. During this process, the endpoint relations, which are independent of the methods used to calculate the form factors, have been utilized to help us reduce the numbers of the expansion coefficients $a^f_{0,1}$ and constrain the behaviors of the $q^2$-dependence in the low-recoil region for some form factors.

Finally, we obtained  the prediction of the differential branching fraction $d\mathcal{B}/dq^2$, the lepton-side forward-backward asymmetry $A_{FB}$, the longitudinal polarization fractions $F_{L}$ and the CP-averaged normalized angular observable $S_{1cc}$ of the baryonic decay $\Lambda_b\rightarrow \Lambda(1520)l^+l^-$. The behaviors of the $q^2$-dependence of the lepton-side forward-backward asymmetry $A_{FB}$, the longitudinal polarization fractions $F_{L}$ and the CP-averaged normalized angular observable $S_{1cc}$ are similar with that of other approaches in most of the $q^2$ region. However, our results for the lepton-side forward-backward asymmetry $A_{FB}$ in the low-recoil region don't show second zero-crossing, which is shown in the prediction of the LQCD \cite{Meinel:2020owd,Meinel:2021mdj}, the NRQM \cite{Mott:2015zma}and the dispersive analysis \cite{Amhis:2022vcd}. Therefore, we hope that more future experimental measurements can give us more meaningful information about these observables.  We also obtained the prediction of the differential branching fraction $d\mathcal{B}/dq^2$, especially for the $\mu$ channel. Until now, compared with other theoretical predictions, only our predictions are the most consistent with the current experimental measurement in the whole $q^2$ region. Moreover, we only performed the tree-level calculation, and the next leading corrections in  both $\alpha_{s}$ and $\Lambda_{QCD}/m_b$ to the results on the partonic expressions of the correlation function can be taken into account to increase the accuracy in the future.

\section{ACknowledgements}
The authors are grateful to Jibo He for the useful discussions.
This work is supported in part by Natural Science Foundation of China under
grant No. 12335003 and 12405114, and by the Fundamental Research Funds for the Central Universities under No. lzujbky-2024-oy02.
\appendix
\section{The results of the the correlation function on the hadronic representation}\label{sec:Appendix A}

For the vector current $j_{\mu,V}$, we have:
\begin{align}
    \Pi_{\lambda\mu,V}&=\frac{-\lambda_{-}\gamma_5}{m^2_{\Lambda^{*}_{-}}-p^2}\Bigg\{f^{V}_{g-}(m_{\Lambda_{b}}+m_{\Lambda^{*}_{-}})g_{\lambda\mu}-f^{V}_{g-}g_{\lambda\mu}\not q 
+q_{\lambda}\Bigg(\bigg[f^{V}_{0-}\frac{2(m_{\Lambda_{b}}+m_{\Lambda^{*}_{-}})^2}{s_{+}}\nonumber \\
&-f^{V}_{\perp-}\frac{2q^2}{s_{+}}
+f^{V}_{g-}\big(\frac{m_{\Lambda^{*}_{-}}}{s_{-}}(2+\frac{2(m_{\Lambda_{b}}+m_{\Lambda^{*}_{-}})(m_{\Lambda^{*}_{-}}(m_{\Lambda_{b}}+m_{\Lambda^{*}_{-}})-s_{+})}{m_{\Lambda^{*}_{-}}s_{+}}\big)\bigg]p_{\mu} \nonumber \\
&-\bigg[f^{V}_{0-}\frac{2(m_{\Lambda_{b}}+m_{\Lambda^{*}_{-}})}{s_{+}}-f^{V}_{\perp-}\frac{2(m_{\Lambda_{b}}+m_{\Lambda^{*}_{-}})}{s_{+}}+f^{V}_{g-}\frac{m_{\Lambda^{*}_{-}}}{s_{-}}\frac{2(m_{\Lambda^{*}_{-}}(m_{\Lambda_{b}}+m_{\Lambda^{*}_{-}})-s_{+})}{m_{\Lambda^{*}_{-}}s_{+}}\bigg]p_{\mu}\not q \nonumber \\
&+\bigg[(m_{\Lambda^{*}_{-}}-m_{\Lambda_{b}})(f^V_{\perp-}+f^V_{g-}\frac{m_{\Lambda^{*}_{-}}}{s_{-}})\bigg]\gamma_{\mu}+\bigg[f^V_{\perp-}+f^V_{g-}\frac{m_{\Lambda^{*}_{-}}}{s_{-}}\bigg]\gamma_{\mu}\not q+ \bigg[f^{V}_{t-}\frac{m_{\Lambda_{b}}^2-m_{\Lambda^{*}_{-}}^2}{q^2}\nonumber\\
&+f^{V}_{0-}\frac{(m_{\Lambda_{b}}+m_{\Lambda^{*}_{-}})^2}{s_{+}}\frac{q^2-m_{\Lambda_{b}}^2+m_{\Lambda^{*}_{-}}^2}{q^2}-f^{V}_{\perp-}\frac{2m_{\Lambda^{*}_{-}}(m_{\Lambda_{b}}+m_{\Lambda^{*}_{-}})}{s_{+}}+f^{V}_{g-}\frac{2m_{\Lambda^{*}_{-}}^2(m_{\Lambda_{b}}+m_{\Lambda^{*}_{-}})}{s_{-}s_{+}}\bigg]q_{\mu} \nonumber\\
&-\bigg[f^{V}_{t-}\frac{m_{\Lambda_{b}}-m_{\Lambda^{*}_{-}}}{q^2}+f^{V}_{0-}\frac{m_{\Lambda_{b}}+m_{\Lambda^{*}_{-}}}{s_{+}}\frac{q^2-m_{\Lambda_{b}}^2+m_{\Lambda^{*}_{-}}^2}{q^2}-f^{V}_{\perp-}\frac{2m_{\Lambda^{*}_{-}}}{s_{+}}+f^{V}_{g-}\frac{2m_{\Lambda^{*}_{-}}^2}{s_{-}s_{+}}\bigg]q_{\mu}\not q\Bigg )\Bigg \} \nonumber\\
&+\frac{-\lambda_{+}\gamma_5}{m^2_{\Lambda^{*}_{+}}-p^2}\Bigg\{f^{V}_{g+}(m_{\Lambda^{*}_{+}}-m_{\Lambda_{b}})g_{\lambda\mu}+f^{V}_{g+}g_{\lambda\mu}\not q 
+q_{\lambda}\Bigg(\bigg[-f^{V}_{0+}\frac{2(m_{\Lambda_{b}}-m_{\Lambda^{*}_{+}})^2}{s_{-}}\nonumber \\
&+f^{V}_{\perp+}\frac{2q^2}{s_{-}}
-f^{V}_{g+}\big(\frac{m_{\Lambda^{*}_{+}}}{s_{+}}(\frac{2(m_{\Lambda^{*}_{+}}-m_{\Lambda_{b}})(m_{\Lambda^{*}_{+}}(m_{\Lambda_{b}}-m_{\Lambda^{*}_{+}})+s_{-})}{m_{\Lambda^{*}_{+}}s_{-}}-2\big)\bigg]p_{\mu} \nonumber \\
&+\bigg[f^{V}_{0+}\frac{2(m_{\Lambda_{b}}-m_{\Lambda^{*}_{+}})}{s_{-}}-f^{V}_{\perp+}\frac{2(m_{\Lambda_{b}}-m_{\Lambda^{*}_{+}})}{s_{-}}-f^{V}_{g+}\frac{m_{\Lambda^{*}_{+}}}{s_{+}}\frac{2(m_{\Lambda^{*}_{+}}(m_{\Lambda_{b}}-m_{\Lambda^{*}_{-}})+s_{-})}{m_{\Lambda^{*}_{+}}s_{-}}\bigg]p_{\mu}\not q \nonumber \\
&+\bigg[(m_{\Lambda^{*}_{+}}+m_{\Lambda_{b}})(f^V_{\perp+}-f^V_{g+}\frac{m_{\Lambda^{*}_{+}}}{s_{+}})\bigg]\gamma_{\mu}-\bigg[f^V_{\perp+}-f^V_{g+}\frac{m_{\Lambda^{*}_{+}}}{s_{+}}\bigg]\gamma_{\mu}\not q-\bigg[f^{V}_{t+}\frac{m_{\Lambda_{b}}^2-m_{\Lambda^{*}_{-}}^2}{q^2}\nonumber\\
&+f^{V}_{0+}\frac{(m_{\Lambda_{b}}-m_{\Lambda^{*}_{+}})^2}{s_{-}}\frac{q^2-m_{\Lambda_{b}}^2+m_{\Lambda^{*}_{-}}^2}{q^2}+f^{V}_{\perp+}\frac{2m_{\Lambda^{*}_{+}}(m_{\Lambda_{b}}-m_{\Lambda^{*}_{+}})}{s_{-}}+f^{V}_{g+}\frac{2m_{\Lambda^{*}_{+}}^2(m_{\Lambda_{b}}-m_{\Lambda^{*}_{+}})}{s_{-}s_{+}}\bigg]q_{\mu} \nonumber\\
&+\bigg[f^{V}_{t+}\frac{m_{\Lambda_{b}}+m_{\Lambda^{*}_{+}}}{q^2}+f^{V}_{0+}\frac{m_{\Lambda_{b}}-m_{\Lambda^{*}_{+}}}{s_{-}}\frac{q^2-m_{\Lambda_{b}}^2+m_{\Lambda^{*}_{+}}^2}{q^2}+f^{V}_{\perp+}\frac{2m_{\Lambda^{*}_{+}}}{s_{-}}+f^{V}_{g+}\frac{2m_{\Lambda^{*}_{+}}^2}{s_{-}s_{+}}\bigg]q_{\mu}\not q\Bigg )\Bigg \} \nonumber \\
&+\int ^{\infty}_{s^{h}_{0}}\frac{ds}{s-p^{2}}\Bigg(\rho^{V}_{1}(s,q^2)g_{\lambda\mu}+\rho^{V}_{2}(s,q^2)g_{\lambda\mu}\not q +q_{\lambda}\bigg[\rho^{V}_{3}(s,q^2)p_{\mu}\nonumber \\
&+\rho^{V}_{4}(s,q^2)p_{\mu}\not q+\rho^{V}_{5}(s,q^2)\gamma_{\mu}+\rho^{V}_{6}(s,q^2)\gamma_{\mu}\not q+\rho^{V}_{7}(s,q^2)q_{\mu}+\rho^{V}_{8}(s,q^2)q_{\mu}\not q\bigg]\Bigg)\ ,
\end{align}
For the axial-vector current $j_{\mu,A}$, we have:
\begin{align}
    \Pi_{\lambda\mu,A}=-\lambda_{5}\times\Pi_{\lambda\mu,V}(\lambda_{-}\rightarrow \lambda_{+}, \ \ \lambda_{+}\rightarrow \lambda_{-}, \ \ f^{V}_{i,-}\rightarrow g^{A}_{i,+},\ \  f^{V}_{i,+}\rightarrow g^{A}_{i,-},\ \ \rho^{V}_{i}\rightarrow\rho^{A}_{i})\ ,
\end{align}
For the tensor current $j_{\mu,T}$, we have:
\begin{align}
    \Pi_{\lambda\mu,T}&=\frac{\lambda_{-}\gamma_5}{m^2_{\Lambda^{*}_{-}}-p^2}\Bigg\{f^{T}_{g-}(m_{\Lambda_{b}}+m_{\Lambda^{*}_{-}})g_{\lambda\mu}-f^{T}_{g-}g_{\lambda\mu}\not q 
+q_{\lambda}\Bigg(\bigg[f^{T}_{0-}\frac{2(m_{\Lambda_{b}}+m_{\Lambda^{*}_{-}})q^2}{s_{+}}\nonumber \\
&-f^{T}_{\perp-}\frac{2(m_{\Lambda_{b}}+m_{\Lambda^{*}_{-}})q^2}{s_{+}}
+f^{T}_{g-}\big(\frac{m_{\Lambda^{*}_{-}}}{s_{-}}(2+\frac{2(m_{\Lambda_{b}}+m_{\Lambda^{*}_{-}})(m_{\Lambda^{*}_{-}}(m_{\Lambda_{b}}+m_{\Lambda^{*}_{-}})-s_{+})}{m_{\Lambda^{*}_{-}}s_{+}}\big)\bigg]p_{\mu} \nonumber \\
&-\bigg[f^{T}_{0-}\frac{2q^{2}}{s_{+}}-f^{T}_{\perp-}\frac{2(m_{\Lambda_{b}}+m_{\Lambda^{*}_{-}})^{2}}{s_{+}}+f^{T}_{g-}\frac{m_{\Lambda^{*}_{-}}}{s_{-}}\frac{2(m_{\Lambda^{*}_{-}}(m_{\Lambda_{b}}+m_{\Lambda^{*}_{-}})-s_{+})}{m_{\Lambda^{*}_{-}}s_{+}}\bigg]p_{\mu}\not q \nonumber \\
&+\bigg[(m_{\Lambda^{*}_{-}}-m_{\Lambda_{b}})(f^{T}_{\perp-}(m_{\Lambda^{*}_{-}}+m_{\Lambda_{b}})+f^{T}_{g-}\frac{m_{\Lambda^{*}_{-}}}{s_{-}})\bigg]\gamma_{\mu}+\bigg[f^{T}_{\perp-}(m_{\Lambda^{*}_{-}}+m_{\Lambda_{b}})+f^V_{g-}\frac{m_{\Lambda^{*}_{-}}}{s_{-}}\bigg]\gamma_{\mu}\not q \nonumber\\
&+\bigg[f^{T}_{0-}\frac{(q^2-m_{\Lambda_{b}}^2+m_{\Lambda^{*}_{-}}^2)(m_{\Lambda_{b}}+m_{\Lambda^{*}_{-}})}{s_{+}}-f^{T}_{\perp-}\frac{2m_{\Lambda^{*}_{-}}(m_{\Lambda_{b}}+m_{\Lambda^{*}_{-}})^{2}}{s_{+}}+f^{T}_{g-}\frac{2m_{\Lambda^{*}_{-}}^2(m_{\Lambda_{b}}+m_{\Lambda^{*}_{-}})}{s_{-}s_{+}}\bigg]q_{\mu} \nonumber\\
&-\bigg[f^{T}_{0-}\frac{q^2-m_{\Lambda_{b}}^2+m_{\Lambda^{*}_{-}}^2}{s_{+}}-f^{T}_{\perp-}\frac{2m_{\Lambda^{*}_{-}}(m_{\Lambda^{*}_{-}}+m_{\Lambda_{b}})}{s_{+}}+f^{V}_{g-}\frac{2m_{\Lambda^{*}_{-}}^2}{s_{-}s_{+}}\bigg]q_{\mu}\not q\Bigg )\Bigg \} \nonumber\\
&+\frac{\lambda_{+}\gamma_5}{m^2_{\Lambda^{*}_{+}}-p^2}\Bigg\{f^{T}_{g+}(m_{\Lambda^{*}_{+}}-m_{\Lambda_{b}})g_{\lambda\mu}+f^{T}_{g+}g_{\lambda\mu}\not q 
+q_{\lambda}\Bigg(\bigg[-f^{T}_{0+}\frac{2(m_{\Lambda_{b}}-m_{\Lambda^{*}_{+}})q^2}{s_{-}}\nonumber \\
&+f^{T}_{\perp+}\frac{2q^2(m_{\Lambda_{b}}-m_{\Lambda^{*}_{+}})}{s_{-}}
-f^{T}_{g+}\big(\frac{m_{\Lambda^{*}_{+}}}{s_{+}}(\frac{2(m_{\Lambda^{*}_{+}}-m_{\Lambda_{b}})(m_{\Lambda^{*}_{+}}(m_{\Lambda_{b}}-m_{\Lambda^{*}_{+}})+s_{-})}{m_{\Lambda^{*}_{+}}s_{-}}-2\big)\bigg]p_{\mu} \nonumber \\
&+\bigg[f^{T}_{0+}\frac{2q^{2}}{s_{-}}-f^{T}_{\perp+}\frac{2(m_{\Lambda_{b}}-m_{\Lambda^{*}_{+}})^{2}}{s_{-}}-f^{V}_{g+}\frac{m_{\Lambda^{*}_{+}}}{s_{+}}\frac{2(m_{\Lambda^{*}_{+}}(m_{\Lambda_{b}}-m_{\Lambda^{*}_{-}})+s_{-})}{m_{\Lambda^{*}_{+}}s_{-}}\bigg]p_{\mu}\not q \nonumber \\
&+\bigg[(m_{\Lambda^{*}_{+}}+m_{\Lambda_{b}})(f^{T}_{\perp+}(m_{\Lambda_{b}}-m_{\Lambda^{*}_{+}})-f^{T}_{g+}\frac{m_{\Lambda^{*}_{+}}}{s_{+}})\bigg]\gamma_{\mu}-\bigg[f^{T}_{\perp+}(m_{\Lambda_{b}}-m_{\Lambda^{*}_{+}})-f^{T}_{g+}\frac{m_{\Lambda^{*}_{+}}}{s_{+}}\bigg]\gamma_{\mu}\not q\nonumber\\
&-\bigg[f^{T}_{0+}\frac{(m_{\Lambda_{b}}-m_{\Lambda^{*}_{+}})(q^2-m_{\Lambda_{b}}^2+m_{\Lambda^{*}_{-}}^2)}{s_{-}}+f^{T}_{\perp+}\frac{2m_{\Lambda^{*}_{+}}(m_{\Lambda_{b}}-m_{\Lambda^{*}_{+}})^{2}}{s_{-}}+f^{T}_{g+}\frac{2m_{\Lambda^{*}_{+}}^2(m_{\Lambda_{b}}-m_{\Lambda^{*}_{+}})}{s_{-}s_{+}}\bigg]q_{\mu} \nonumber\\
&+\bigg[f^{T}_{0+}\frac{q^2-m_{\Lambda_{b}}^2+m_{\Lambda^{*}_{+}}^2}{s_{-}}+f^{T}_{\perp+}\frac{2m_{\Lambda^{*}_{+}}(m_{\Lambda_{b}}-m_{\Lambda^{*}_{+}})}{s_{-}}+f^{T}_{g+}\frac{2m_{\Lambda^{*}_{+}}^2}{s_{-}s_{+}}\bigg]q_{\mu}\not q\Bigg )\Bigg \}\nonumber\\
&+\int ^{\infty}_{s^{h}_{0}}\frac{ds}{s-p^{2}}\Bigg(\rho^{T}_{1}(s,q^2)g_{\lambda\mu}+\rho^{T}_{2}(s,q^2)g_{\lambda\mu}\not q +q_{\lambda}\bigg[\rho^{T}_{3}(s,q^2)p_{\mu}\nonumber \\
&+\rho^{T}_{4}(s,q^2)p_{\mu}\not q+\rho^{T}_{5}(s,q^2)\gamma_{\mu}+\rho^{T}_{6}(s,q^2)\gamma_{\mu}\not q+\rho^{T}_{7}(s,q^2)q_{\mu}+\rho^{T}_{8}(s,q^2)q_{\mu}\not q\bigg]\Bigg)\ ,
\end{align}
For the axial-tensor current $j_{\mu,T5}$, we have:
\begin{align}
    \Pi_{\lambda\mu,T}=\lambda_{5}\times\Pi_{\lambda\mu,T}(\lambda_{-}\rightarrow \lambda_{+}, \ \ \lambda_{+}\rightarrow \lambda_{-}, \ \ f^{T}_{i,-}\rightarrow g^{T5}_{i,+},\ \  f^{T}_{i,+}\rightarrow g^{T5}_{i,-},\ \ \rho^{T}_{i}\rightarrow\rho^{T5}_{i})\ .
\end{align}

\section{The results of  the transformed coefficients $\rho^{d}_{i,n}(s,q^2)$ on the partonic representation}\label{sec:Appendix B}
The coefficients of each Lorentz structure on the partonic representation for each weak transition current $j_{\mu,d}$ are given in form of Eq.\ref{EQ:invaramp}, where the transformed coefficient functions $\rho^{d}_{i,n}(s,q^2)$ with  d=V, A, T, T5(the Lorentz structure of the weak transition current $\bar{s}\Gamma_{\mu,d}b$), $i=1,..,8$(the number of the coefficient of each Lorentz structure on the partonic representation) and $n=1,2$(the power of the denominator D) are listed below:  
\begin{itemize}
\item For the vector current $j_{\mu,V}$:
\begin{eqnarray}
\rho^{\rm V}_{1,n}&&=\rho^{\rm V}_{2,n}(n=1-2)=0\ , \ \ \rho^{V}_{3,2}=\rho^{V}_{4,2}=\rho^{V}_{8,2}=0\ ,\nonumber\\
\rho^{\rm V}_{3,1}&&=\frac{2}{\sqrt{6}}\times \big[8\sigma\bar{\sigma}f^{(2)}_{\Lambda}(\overline{\psi_{4}}-\overline{\psi_{2}})+\frac{8(q^{2}-m^{2}_{\Lambda_{b}}\bar{\sigma}^{2}-m^{2}_{s})}{m^{2}_{\Lambda_{b}}}f^{(2)}_{\Lambda}\overline{\psi_{2}}-\frac{16m_{s}}{m_{\Lambda_{b}}}f^{(1)}_{\Lambda}\overline{\psi^{\sigma}_{3}}\big] \ ,\nonumber \\
\rho^{V}_{4,1}&&=\rho^{V}_{8,1}=\frac{2}{\sqrt{6}}\times \big[-\frac{16\sigma}{m_{\Lambda_{b}}}f^{(1)}_{\Lambda}\overline{\psi^{\sigma}_{3}}\big]\ ,\nonumber \\
\rho^{V}_{5,1}&&=\frac{2}{\sqrt{6}}\times\big[4\sigma (m_{s}-\bar{\sigma}m_{\Lambda_{b}})f^{(2)}_{\Lambda}(\overline{\psi_{4}}-\overline{\psi_{2}})-\frac{8(m^{2}_{s}-\sigma q^{2}-m_{s}m_{\Lambda_{b}}\bar{\sigma})}{m_{\Lambda_{b}}\bar{\sigma}}f^{(1)}_{\Lambda}\overline{\psi^{\sigma}_{3}}\nonumber\\
&&\ \ \ +\frac{4(q^{2}-m^{2}_{\Lambda_{b}}\bar{\sigma}^{2}-m^{2}_{s})(m_{s}-\bar{\sigma}m_{\Lambda_{b}})}{m^{2}_{\Lambda_{b}}\bar{\sigma}}f^{(2)}_{\Lambda}\overline{\psi_{2}} \big]\ ,\nonumber \\
\rho^{V}_{5,2}&&=-\frac{2}{\sqrt{6}}\times\big[\frac{4m_{s}}{m^{2}_{\Lambda_{b}}\bar{\sigma}}f^{(2)}_{\Lambda}\overline{\psi_{2}}+\frac{8\sigma}{m_{\Lambda_{b}}\bar{\sigma}}f^{(1)}_{\Lambda}\overline{\psi^{\sigma}_{3}}\big]\ ,  \nonumber \\
\rho^{V}_{6,1}&&=\frac{2}{\sqrt{6}}\times\big[4\sigma f^{(2)}_{\Lambda}(\overline{\psi_{4}}-\overline{\psi_{2}})+\frac{4(q^{2}-m^{2}_{\Lambda_{b}}\bar{\sigma}^{2}-m^{2}_{s})}{m^{2}_{\Lambda_{b}}\bar{\sigma}}f^{(2)}_{\Lambda}\overline{\psi_{2}}-\frac{8(m_{s}+\sigma m_{\Lambda_{b}})}{m_{\Lambda_{b}}}f^{(1)}_{\Lambda}\overline{\psi^{\sigma}_{3}}\big]\ ,\nonumber\\
\rho^{V}_{6,2}&&=-\frac{2}{\sqrt{6}}\times\big[\frac{4}{m^{2}_{\Lambda_{b}}\bar{\sigma}}f^{(2)}_{\Lambda}\overline{\psi_{2}}\big]\ ,\nonumber\\
\rho^{V}_{7,1}&&=\frac{2}{\sqrt{6}}\times\big[16\sigma f^{(1)}_{\Lambda}\overline{\psi^{\sigma}_{3}}
        -8\sigma^{2} f^{(2)}_{\Lambda}(\overline{\psi_{4}}-\overline{\psi_{2}})-\frac{8\sigma(q^{2}-m^{2}_{\Lambda_{b}}\bar{\sigma}^{2}-m^{2}_{s})}{m^{2}_{\Lambda_{b}}\bar{\sigma}}f^{(2)}_{\Lambda}\overline{\psi_{2}}\big]\ ,\nonumber\\
\rho^{V}_{7,2}&&=\frac{2}{\sqrt{6}}\times\big[\frac{8}{m^{2}_{\Lambda_{b}}\bar{\sigma}}f^{(2)}_{\Lambda}\overline{\psi_{2}}\big]\ ,       
\end{eqnarray}
where the functions $\overline{\psi}(\omega,u)$ are defined as:
\begin{eqnarray}
	\overline{\psi}(\omega, u)=\int^{\omega}_0 d\tau \ \tau\psi(\tau,u)\ ,
\end{eqnarray}
originating from the partial integral in the variable $\omega$ in Eq.(\ref{EQ:invaramp})
\item For the axlai-vector current $j_{\mu,A}$:
\begin{eqnarray}
\rho^{\rm A}_{1,n}&&=\rho^{\rm A}_{2,n}(n=1-2)=0\ , \ \ \rho^{A}_{3,2}=\rho^{A}_{4,2}=\rho^{V}_{8,2}=0\ ,\nonumber\\
\rho^{\rm A}_{3,1}&&=\frac{2}{\sqrt{6}}\times \big[8\sigma\bar{\sigma}f^{(2)}_{\Lambda}(\overline{\psi_{4}}-\overline{\psi_{2}})+\frac{8(q^{2}-m^{2}_{\Lambda_{b}}\bar{\sigma}^{2}-m^{2}_{s})}{m^{2}_{\Lambda_{b}}}f^{(2)}_{\Lambda}\overline{\psi_{2}}-\frac{16m_{s}}{m_{\Lambda_{b}}}f^{(1)}_{\Lambda}\overline{\psi^{\sigma}_{3}}\big]\ ,\nonumber \\
\rho^{A}_{4,1}&&=\rho^{A}_{8,1}=\frac{2}{\sqrt{6}}\times \big[\frac{16\sigma}{m_{\Lambda_{b}}}f^{(1)}_{\Lambda}\overline{\psi^{\sigma}_{3}}\big]\ ,\nonumber \\
\rho^{A}_{5,1}&&=\frac{2}{\sqrt{6}}\times\big[-4\sigma (m_{s}+\bar{\sigma}m_{\Lambda_{b}})f^{(2)}_{\Lambda}(\overline{\psi_{4}}-\overline{\psi_{2}})+\frac{8(m^{2}_{s}-\sigma q^{2}+m_{s}m_{\Lambda_{b}}\bar{\sigma})}{m_{\Lambda_{b}}\bar{\sigma}}f^{(1)}_{\Lambda}\overline{\psi^{\sigma}_{3}}\nonumber\\
&&\ \ \ -\frac{4(q^{2}-m^{2}_{\Lambda_{b}}\bar{\sigma}^{2}-m^{2}_{s})(m_{s}+\bar{\sigma}m_{\Lambda_{b}})}{m^{2}_{\Lambda_{b}}\bar{\sigma}}f^{(2)}_{\Lambda}\overline{\psi_{2}} \big]\ ,\nonumber \\
\rho^{A}_{5,2}&&=\frac{2}{\sqrt{6}}\times\big[\frac{4m_{s}}{m^{2}_{\Lambda_{b}}\bar{\sigma}}f^{(2)}_{\Lambda}\overline{\psi_{2}}+\frac{8\sigma}{m_{\Lambda_{b}}\bar{\sigma}}f^{(1)}_{\Lambda}\overline{\psi^{\sigma}_{3}}\big]\ ,\nonumber \\
\rho^{A}_{6,1}&&=\frac{2}{\sqrt{6}}\times\big[4\sigma f^{(2)}_{\Lambda}(\overline{\psi_{4}}-\overline{\psi_{2}})+\frac{4(q^{2}-m^{2}_{\Lambda_{b}}\bar{\sigma}^{2}-m^{2}_{s})}{m^{2}_{\Lambda_{b}}\bar{\sigma}}f^{(2)}_{\Lambda}\overline{\psi_{2}}-\frac{8(m_{s}-\sigma m_{\Lambda_{b}})}{m_{\Lambda_{b}}}f^{(1)}_{\Lambda}\overline{\psi^{\sigma}_{3}}\big]\ ,\nonumber\\
\rho^{A}_{6,2}&&=-\frac{2}{\sqrt{6}}\times\big[\frac{4}{m^{2}_{\Lambda_{b}}\bar{\sigma}}f^{(2)}_{\Lambda}\overline{\psi_{2}}\big]\ ,\nonumber\\
\rho^{A}_{7,1}&&=\frac{2}{\sqrt{6}}\times\big[-16\sigma f^{(1)}_{\Lambda}\overline{\psi^{\sigma}_{3}}
        -8\sigma^{2} f^{(2)}_{\Lambda}(\overline{\psi_{4}}-\overline{\psi_{2}})-\frac{8\sigma(q^{2}-m^{2}_{\Lambda_{b}}\bar{\sigma}^{2}-m^{2}_{s})}{m^{2}_{\Lambda_{b}}\bar{\sigma}}f^{(2)}_{\Lambda}\overline{\psi_{2}}\big]\ ,\nonumber\\
\rho^{A}_{7,2}&&=\frac{2}{\sqrt{6}}\times\big[\frac{8}{m^{2}_{\Lambda_{b}}\bar{\sigma}}f^{(2)}_{\Lambda}\overline{\psi_{2}}\big] \ ,      
\end{eqnarray}
\item For the tensor current $j_{\mu,T}$:
\begin{eqnarray}
\rho^{\rm T}_{1,n}&&=\rho^{\rm T}_{2,n}(n=1-2)=0\ , \ \ \rho^{T}_{3,2}=\rho^{T}_{4,2}=0\ ,\nonumber\\
\rho^{\rm T}_{3,1}&&=\frac{2}{\sqrt{6}}\times \big[\frac{16\sigma q^{2}}{m_{\Lambda_{b}}}f^{(1)}_{\Lambda}\overline{\psi^{\sigma}_{3}}\big]\ ,\nonumber \\
\rho^{T}_{4,1}&&=\frac{2}{\sqrt{6}}\times \big[\frac{16m_{s}}{m_{\Lambda_{b}}}f^{(1)}_{\Lambda}\overline{\psi^{\sigma}_{3}}-\frac{8(q^{2}-m^{2}_{\Lambda_{b}}\bar{\sigma}^{2}-m^{2}_{s})}{m^{2}_{\Lambda_{b}}}f^{(2)}_{\Lambda}\overline{\psi_{2}}-8\sigma\bar{\sigma}f^{(2)}_{\Lambda}(\overline{\psi_{4}}-\overline{\psi_{2}})\big]\ ,\nonumber \\
\rho^{T}_{5,1}&&=\frac{2}{\sqrt{6}}\times\big[\frac{4(q^{2}-m^{2}_{\Lambda_{b}}\bar{\sigma}^{2}-m^{2}_{s})(\bar{\sigma}^{2}m^{2}_{\Lambda_{b}}-m^{2}_{s})}{m^{2}_{\Lambda_{b}}\bar{\sigma}}f^{(2)}_{\Lambda}\overline{\psi_{2}}+4\sigma(\bar{\sigma}^{2}m^{2}_{\Lambda_{b}}-m^{2}_{s})f^{(2)}_{\Lambda}(\overline{\psi_{4}}-\overline{\psi_{2}})\nonumber \\
&&\ \ \ +\frac{8(m^{3}_{s}-\sigma q^{2}(\bar{\sigma}m_{\Lambda_{b}}+m_{s})-\bar{\sigma}^{2}m^{2}_{\Lambda_{b}}m_{s})}{m_{\Lambda_{b}}\bar{\sigma}}f^{(1)}_{\Lambda}\overline{\psi^{\sigma}_{3}}\big]\ ,\nonumber \\
\rho^{T}_{5,2}&&=\frac{2}{\sqrt{6}}\times\big[\frac{4(\bar{\sigma}^{2}m^{2}_{\Lambda_{b}}+m^{2}_{s})}{m^{2}_{\Lambda_{b}}\bar{\sigma}}f^{(2)}_{\Lambda}\overline{\psi_{2}}+\frac{8m_{s}}{m_{\Lambda_{b}}\bar{\sigma}}f^{(1)}_{\Lambda}\overline{\psi^{\sigma}_{3}}-4\sigma f^{(2)}_{\Lambda}(\overline{\psi_{4}}-\overline{\psi_{2}})\big]\ , \nonumber \\
\rho^{T}_{6,1}&&=\frac{2}{\sqrt{6}}\times\big[\frac{8(m^{2}_{s}+\sigma\bar{\sigma}m^{2}_{\Lambda_{b}}+m_{\Lambda_{b}}m_{s})}{m_{\Lambda_{b}}}f^{(1)}_{\Lambda}\overline{\psi^{\sigma}_{3}}-4\sigma(\bar{\sigma}m_{\Lambda_{b}}+m_{s})f^{(2)}_{\Lambda}(\overline{\psi_{4}}-\overline{\psi_{2}})\nonumber\\
&&\ \ \ -\frac{4(q^{2}-m^{2}_{\Lambda_{b}}\bar{\sigma}^{2}-m^{2}_{s})(\bar{\sigma}m_{\Lambda_{b}}+m_{s})}{m^{2}_{\Lambda_{b}}\bar{\sigma}}f^{(2)}_{\Lambda}\overline{\psi_{2}}\big]\ ,\nonumber\\
\rho^{T}_{6,2}&&=\frac{2}{\sqrt{6}}\times\big[\frac{4m_{s}}{m^{2}_{\Lambda_{b}}\bar{\sigma}}f^{(2)}_{\Lambda}\overline{\psi_{2}}\big]\ ,\nonumber\\
\rho^{T}_{7,1}&&=\frac{2}{\sqrt{6}}\times\big[\frac{4(q^{2}-m^{2}_{\Lambda_{b}}\bar{\sigma}^{2}-m^{2}_{s})(\bar{\sigma}m_{\Lambda_{b}}+m_{s})}{m^{2}_{\Lambda_{b}}\bar{\sigma}}f^{(2)}_{\Lambda}\overline{\psi_{2}}+4\sigma(\bar{\sigma}m_{\Lambda_{b}}+m_{s})f^{(2)}_{\Lambda}(\overline{\psi_{4}}-\overline{\psi_{2}})\nonumber\\
&&\ \ \ +\frac{8((2\sigma -1)m^{2}_{s}-2\sigma\bar{\sigma}^{2}m^{2}_{\Lambda_{b}}-\sigma q^{2}(2\sigma-1)-\bar{\sigma}m_{\Lambda_{b}}m_{s})}{m_{\Lambda_{b}}\bar{\sigma}}f^{(1)}_{\Lambda}\overline{\psi^{\sigma}_{3}}\big]\ ,\nonumber\\
\rho^{T}_{7,2}&&=\frac{2}{\sqrt{6}}\times\big[\frac{8\sigma}{m_{\Lambda_{b}}\bar{\sigma}}f^{(1)}_{\Lambda}\overline{\psi^{\sigma}_{3}}-\frac{4m_{s}}{m^{2}_{\Lambda_{b}}\bar{\sigma}}f^{(2)}_{\Lambda}\overline{\psi_{2}}\big]\ ,\nonumber\\
\rho^{T}_{8,1}&&=\frac{2}{\sqrt{6}}\times\big[\frac{8(m_{s}+m_{\Lambda_{b}}\sigma)}{m_{\Lambda_{b}}}f^{(1)}_{\Lambda}\overline{\psi^{\sigma}_{3}}+4\sigma(2\sigma -1)f^{(2)}_{\Lambda}(\overline{\psi_{4}}-\overline{\psi_{2}})\nonumber\\
&&\ \ \ +\frac{4(q^{2}-m^{2}_{\Lambda_{b}}\bar{\sigma}^{2}-m^{2}_{s})(2\sigma-1)}{m^{2}_{\Lambda_{b}}\bar{\sigma}}f^{(2)}_{\Lambda}\overline{\psi_{2}}\big]\ ,\nonumber\\
\rho^{T}_{8,2}&&=-\frac{2}{\sqrt{6}}\times\big[\frac{4}{m^{2}_{\Lambda_{b}}\bar{\sigma}}f^{(2)}_{\Lambda}\overline{\psi_{2}}\big]\ ,
\end{eqnarray}
\item For the axial-tensor current $j_{\mu,T5}$:
\begin{eqnarray}
\rho^{\rm T5}_{1,n}&&=\rho^{\rm T5}_{2,n}(n=1-2)=0\ , \ \ \rho^{T5}_{3,2}=\rho^{T5}_{4,2}=0\ ,\nonumber\\
\rho^{\rm T5}_{3,1}&&=\frac{2}{\sqrt{6}}\times \big[\frac{16\sigma q^{2}}{m_{\Lambda_{b}}}f^{(1)}_{\Lambda}\overline{\psi^{\sigma}_{3}}\big]\ ,\nonumber \\
\rho^{T5}_{4,1}&&=\frac{2}{\sqrt{6}}\times \big[-\frac{16m_{s}}{m_{\Lambda_{b}}}f^{(1)}_{\Lambda}\overline{\psi^{\sigma}_{3}}-\frac{8(m^{2}_{\Lambda_{b}}\bar{\sigma}^{2}+m^{2}_{s}-q^{2})}{m^{2}_{\Lambda_{b}}}f^{(2)}_{\Lambda}\overline{\psi_{2}}-8\sigma\bar{\sigma}f^{(2)}_{\Lambda}(\overline{\psi_{2}}-\overline{\psi_{4}})\big]\ ,\nonumber \\
\rho^{T5}_{5,1}&&=\frac{2}{\sqrt{6}}\times\big[\frac{4(m^{2}_{\Lambda_{b}}\bar{\sigma}^{2}+m^{2}_{s}-q^{2})(\bar{\sigma}^{2}m^{2}_{\Lambda_{b}}-m^{2}_{s})}{m^{2}_{\Lambda_{b}}\bar{\sigma}}f^{(2)}_{\Lambda}\overline{\psi_{2}}+4\sigma(\bar{\sigma}^{2}m^{2}_{\Lambda_{b}}-m^{2}_{s})f^{(2)}_{\Lambda}(\overline{\psi_{2}}-\overline{\psi_{4}})\nonumber \\
&&\ \ \ +\frac{8(\sigma q^{2}(m_{s}-\bar{\sigma}m_{\Lambda_{b}})+\bar{\sigma}^{2}m^{2}_{\Lambda_{b}}m_{s}-m^{3}_{s})}{m_{\Lambda_{b}}\bar{\sigma}}f^{(1)}_{\Lambda}\overline{\psi^{\sigma}_{3}}\big]\ ,\nonumber \\
\rho^{T5}_{5,2}&&=-\frac{2}{\sqrt{6}}\times\big[\frac{4(\bar{\sigma}^{2}m^{2}_{\Lambda_{b}}+m^{2}_{s})}{m^{2}_{\Lambda_{b}}\bar{\sigma}}f^{(2)}_{\Lambda}\overline{\psi_{2}}+\frac{8m_{s}}{m_{\Lambda_{b}}\bar{\sigma}}f^{(1)}_{\Lambda}\overline{\psi^{\sigma}_{3}}+4\sigma f^{(2)}_{\Lambda}(\overline{\psi_{2}}-\overline{\psi_{4}})\big]\ , \nonumber \\
\rho^{T5}_{6,1}&&=\frac{2}{\sqrt{6}}\times\big[\frac{8(m^{2}_{s}+\sigma\bar{\sigma}m^{2}_{\Lambda_{b}}-m_{\Lambda_{b}}m_{s})}{m_{\Lambda_{b}}}f^{(1)}_{\Lambda}\overline{\psi^{\sigma}_{3}}-4\sigma(\bar{\sigma}m_{\Lambda_{b}}-m_{s})f^{(2)}_{\Lambda}(\overline{\psi_{2}}-\overline{\psi_{4}})\nonumber\\
&&\ \ \ -\frac{4(m^{2}_{\Lambda_{b}}\bar{\sigma}^{2}+m^{2}_{s}-q^{2})(\bar{\sigma}m_{\Lambda_{b}}-m_{s})}{m^{2}_{\Lambda_{b}}\bar{\sigma}}f^{(2)}_{\Lambda}\overline{\psi_{2}}\big]\ ,\nonumber\\
\rho^{T5}_{6,2}&&=\frac{2}{\sqrt{6}}\times\big[\frac{4m_{s}}{m^{2}_{\Lambda_{b}}\bar{\sigma}}f^{(2)}_{\Lambda}\overline{\psi_{2}}\big]\ ,\nonumber\\
\rho^{T5}_{7,1}&&=\frac{2}{\sqrt{6}}\times\big[\frac{4(m^{2}_{\Lambda_{b}}\bar{\sigma}^{2}+m^{2}_{s}-q^{2})(\bar{\sigma}m_{\Lambda_{b}}-m_{s})}{m^{2}_{\Lambda_{b}}\bar{\sigma}}f^{(2)}_{\Lambda}\overline{\psi_{2}}+4\sigma(\bar{\sigma}m_{\Lambda_{b}}-m_{s})f^{(2)}_{\Lambda}(\overline{\psi_{2}}-\overline{\psi_{4}})\nonumber\\
&&\ \ \ +\frac{8((2\sigma -1)m^{2}_{s}-2\sigma\bar{\sigma}^{2}m^{2}_{\Lambda_{b}}-\sigma q^{2}(2\sigma-1)+\bar{\sigma}m_{\Lambda_{b}}m_{s})}{m_{\Lambda_{b}}\bar{\sigma}}f^{(1)}_{\Lambda}\overline{\psi^{\sigma}_{3}}\big]\ ,\nonumber\\
\rho^{T5}_{7,2}&&=\frac{2}{\sqrt{6}}\times\big[\frac{8\sigma}{m_{\Lambda_{b}}\bar{\sigma}}f^{(1)}_{\Lambda}\overline{\psi^{\sigma}_{3}}-\frac{4m_{s}}{m^{2}_{\Lambda_{b}}\bar{\sigma}}f^{(2)}_{\Lambda}\overline{\psi_{2}}\big]\ ,\nonumber\\
\rho^{T5}_{8,1}&&=\frac{2}{\sqrt{6}}\times\big[\frac{8(m_{\Lambda_{b}}\sigma-m_{s})}{m_{\Lambda_{b}}}f^{(1)}_{\Lambda}\overline{\psi^{\sigma}_{3}}+4\sigma(2\sigma -1)f^{(2)}_{\Lambda}(\overline{\psi_{2}}-\overline{\psi_{4}})\nonumber\\
&&\ \ \ +\frac{4(m^{2}_{\Lambda_{b}}\bar{\sigma}^{2}+m^{2}_{s}-q^{2})(2\sigma-1)}{m^{2}_{\Lambda_{b}}\bar{\sigma}}f^{(2)}_{\Lambda}\overline{\psi_{2}}\big]\ ,\nonumber\\
\rho^{T5}_{8,2}&&=\frac{2}{\sqrt{6}}\times\big[\frac{4}{m^{2}_{\Lambda_{b}}\bar{\sigma}}f^{(2)}_{\Lambda}\overline{\psi_{2}}\big].
\end{eqnarray}  
\end{itemize}

\section{The results of  the form factors for $\Lambda_{b}\rightarrow \Lambda(1520)$}\label{sec:Appendix C}

 The final results of all form factors for $\Lambda_{b}\rightarrow \Lambda(1520)$ by using LCSR are listed below:
\begin{eqnarray}
\label{equ::FFs-Vector}
    f^{V}_{t}(q^2)&&=\frac{e^{m_{\Lambda^{*}_{-}}^2/M^{2}}}{\lambda_{-}(m_{\Lambda^{*}_{-}}+m_{\Lambda^{*}_{-}})}\Big[\frac{q^2}{m_{\Lambda_{b}}-m_{\Lambda^{*}_{-}}}\Big(\Pi^{V,B}_{7}+(m_{\Lambda_{b}}-m_{\Lambda^{*}_{+}})\Pi^{V,B}_{8}\nonumber\\
    &&\ \ \ \ +\frac{2m_{\Lambda^{*}_{-}}}{(m_{\Lambda_{b}}+m_{\Lambda^{*}_{-}})^2-q^2}(\Pi^{V,B}_{5}+(m_{\Lambda_{b}}+m_{\Lambda^{*}_{+}})\Pi^{V,B}_{6})\Big)\nonumber\\
    &&\ \ \ -\frac{q^2-m_{\Lambda_{b}}^2+m_{\Lambda^{*}_{-}}^2}{2(m_{\Lambda_{b}}-m_{\Lambda^{*}_{-}})}\Big(\Pi^{V,B}_{3}+(m_{\Lambda_{b}}-m_{\Lambda^{*}_{+}})\Pi^{V,B}_{4}+\frac{2(m_{\Lambda_{b}}+m_{\Lambda^{*}_{-}})}{(m_{\Lambda_{b}}+m_{\Lambda^{*}_{-}})^2-q^2}\Pi^{V,B}_{5}\nonumber\\
    &&\ \ \ \ +\Pi^{V,B}_{6}(\frac{2(m_{\Lambda_{b}}+m_{\Lambda^{*}_{-}})}{(m_{\Lambda_{b}}+m_{\Lambda^{*}_{-}})^2-q^2}(m_{\Lambda_{b}}+m_{\Lambda^{*}_{+}})-2)\Big)\Big]\ ,\nonumber\\
    f^{V}_{0}(q^2)&&=\frac{e^{m_{\Lambda^{*}_{-}}^2/M^{2}}}{2\lambda_{-}(m_{\Lambda^{*}_{-}}+m_{\Lambda^{*}_{-}})} \frac{(m_{\Lambda_{b}}+m_{\Lambda^{*}_{-}})^2-q^2}{m_{\Lambda_{b}}+m_{\Lambda^{*}_{-}}}\Big[\Pi^{V,B}_{3}+(m_{\Lambda_{b}}-m_{\Lambda^{*}_{+}})\Pi^{V,B}_{4}\nonumber\\
    &&\ \ \ +\frac{2(m_{\Lambda_{b}}+m_{\Lambda^{*}_{-}})}{(m_{\Lambda_{b}}+m_{\Lambda^{*}_{-}})^2-q^2}\Pi^{V,B}_{5} +\Pi^{V,B}_{6}(\frac{2(m_{\Lambda_{b}}+m_{\Lambda^{*}_{-}})}{(m_{\Lambda_{b}}+m_{\Lambda^{*}_{-}})^2-q^2}(m_{\Lambda_{b}}+m_{\Lambda^{*}_{+}})-2)\Big]\ , \nonumber\\
    f^{V}_{\perp}(q^2)&&=\frac{e^{m_{\Lambda^{*}_{-}}^2/M^{2}}}{\lambda_{-}(m_{\Lambda^{*}_{-}}+m_{\Lambda^{*}_{-}})}\Big[\Pi^{V,B}_{5}+(m_{\Lambda_{b}}+m_{\Lambda^{*}_{+}})\Pi^{V,B}_{6}\Big]\ ,\nonumber\\
    f^{V}_{g}(q^2)&&=0\ ,
\end{eqnarray}
\begin{eqnarray}
\label{equ::FFs-axial-Vector}
    g^{A}_{t}(q^2)&&=-\frac{e^{m_{\Lambda^{*}_{-}}^2/M^{2}}}{\lambda_{-}(m_{\Lambda^{*}_{-}}+m_{\Lambda^{*}_{-}})}\Big[\frac{q^2}{m_{\Lambda_{b}}+m_{\Lambda^{*}_{-}}}\Big(\Pi^{A,B}_{7}+(m_{\Lambda_{b}}+m_{\Lambda^{*}_{+}})\Pi^{A,B}_{8}\nonumber\\
    &&\ \ \ \ -\frac{2m_{\Lambda^{*}_{-}}}{(m_{\Lambda_{b}}-m_{\Lambda^{*}_{-}})^2-q^2}(\Pi^{A,B}_{5}+(m_{\Lambda_{b}}-m_{\Lambda^{*}_{+}})\Pi^{A,B}_{6})\Big)\nonumber\\
    &&\ \ \ -\frac{q^2-m_{\Lambda_{b}}^2+m_{\Lambda^{*}_{-}}^2}{2(m_{\Lambda_{b}}+m_{\Lambda^{*}_{-}})}\Big(\Pi^{A,B}_{3}+(m_{\Lambda_{b}}+m_{\Lambda^{*}_{+}})\Pi^{A,B}_{4}+\frac{2(m_{\Lambda_{b}}-m_{\Lambda^{*}_{-}})}{(m_{\Lambda_{b}}-m_{\Lambda^{*}_{-}})^2-q^2}\Pi^{A,B}_{5}\nonumber\\
    &&\ \ \ \ +\Pi^{A,B}_{6}(\frac{2(m_{\Lambda_{b}}-m_{\Lambda^{*}_{-}})}{(m_{\Lambda_{b}}-m_{\Lambda^{*}_{-}})^2-q^2}(m_{\Lambda_{b}}-m_{\Lambda^{*}_{+}})-2)\Big)\Big]\ ,\nonumber\\
    g^{A}_{0}(q^2)&&=-\frac{e^{m_{\Lambda^{*}_{-}}^2/M^{2}}}{2\lambda_{-}(m_{\Lambda^{*}_{-}}+m_{\Lambda^{*}_{-}})} \frac{(m_{\Lambda_{b}}-m_{\Lambda^{*}_{-}})^2-q^2}{m_{\Lambda_{b}}-m_{\Lambda^{*}_{-}}}\Big[\Pi^{A,B}_{3}+(m_{\Lambda_{b}}+m_{\Lambda^{*}_{+}})\Pi^{A,B}_{4}\nonumber\\
    &&\ \ \ +\frac{2(m_{\Lambda_{b}}-m_{\Lambda^{*}_{-}})}{(m_{\Lambda_{b}}-m_{\Lambda^{*}_{-}})^2-q^2}\Pi^{A,B}_{5} +\Pi^{A,B}_{6}(\frac{2(m_{\Lambda_{b}}-m_{\Lambda^{*}_{-}})}{(m_{\Lambda_{b}}-m_{\Lambda^{*}_{-}})^2-q^2}(m_{\Lambda_{b}}-m_{\Lambda^{*}_{+}})-2)\Big]\ , \nonumber\\
    g^{A}_{\perp}(q^2)&&=-\frac{e^{m_{\Lambda^{*}_{-}}^2/M^{2}}}{\lambda_{-}(m_{\Lambda^{*}_{-}}+m_{\Lambda^{*}_{-}})}\Big[\Pi^{A,B}_{5}+(m_{\Lambda_{b}}-m_{\Lambda^{*}_{+}})\Pi^{A,B}_{6}\Big]\ ,\nonumber\\
    g^{A}_{g}(q^2)&&=0\ ,
\end{eqnarray}
\begin{eqnarray}
\label{equ::FFs-Tensor-Vector}
    f^{T}_{0}(q^2)&&=-\frac{e^{m_{\Lambda^{*}_{-}}^2/M^{2}}}{\lambda_{-}(m_{\Lambda^{*}_{-}}+m_{\Lambda^{*}_{-}})} \frac{(m_{\Lambda_{b}}+m_{\Lambda^{*}_{-}})^2-q^2}{q^2-m_{\Lambda_{b}}^2+m_{\Lambda^{*}_{-}}^2}\Big[\Pi^{T,B}_{7}+(m_{\Lambda_{b}}-m_{\Lambda^{*}_{+}})\Pi^{T,B}_{8}\nonumber\\
    &&\ \ \ +\frac{2m_{\Lambda^{*}_{-}}}{(m_{\Lambda_{b}}+m_{\Lambda^{*}_{-}})^2-q^2}\Pi^{T,B}_{5} +\Pi^{T,B}_{6}\frac{2m_{\Lambda^{*}_{-}}}{(m_{\Lambda_{b}}+m_{\Lambda^{*}_{-}})^2-q^2}(m_{\Lambda_{b}}+m_{\Lambda^{*}_{+}})\Big]\ , \nonumber\\
    f^{T}_{\perp}(q^2)&&=-\frac{e^{m_{\Lambda^{*}_{-}}^2/M^{2}}}{\lambda_{-}(m_{\Lambda^{*}_{-}}+m_{\Lambda^{*}_{-}})} \frac{1}{(m_{\Lambda_{b}}+m_{\Lambda^{*}_{-}})}\Big[\Pi^{T,B}_{5}+(m_{\Lambda_{b}}+m_{\Lambda^{*}_{+}})\Pi^{T,B}_{6}\Big]\ ,\nonumber\\
    f^{T}_{g-}(q^2)&&=0\ ,
\end{eqnarray}
\begin{eqnarray}
\label{equ::FFs-axial-Tensor-Vector}
    g^{T5}_{0}(q^2)&&=-\frac{e^{m_{\Lambda^{*}_{-}}^2/M^{2}}}{\lambda_{-}(m_{\Lambda^{*}_{-}}+m_{\Lambda^{*}_{-}})} \frac{(m_{\Lambda_{b}}-m_{\Lambda^{*}_{-}})^2-q^2}{q^2-m_{\Lambda_{b}}^2+m_{\Lambda^{*}_{-}}^2}\Big[\Pi^{T5,B}_{7}+(m_{\Lambda_{b}}+m_{\Lambda^{*}_{+}})\Pi^{T5,B}_{8}\nonumber\\
    &&\ \ \ -\frac{2m_{\Lambda^{*}_{-}}}{(m_{\Lambda_{b}}-m_{\Lambda^{*}_{-}})^2-q^2}\Pi^{T5,B}_{5} -\Pi^{T5,B}_{6}\frac{2m_{\Lambda^{*}_{-}}}{(m_{\Lambda_{b}}-m_{\Lambda^{*}_{-}})^2-q^2}(m_{\Lambda_{b}}-m_{\Lambda^{*}_{+}})\Big]\ ,\nonumber\\
    g^{T5}_{\perp}(q^2)&&=-\frac{e^{m_{\Lambda^{*}_{-}}^2/M^{2}}}{\lambda_{-}(m_{\Lambda^{*}_{-}}+m_{\Lambda^{*}_{-}})} \frac{1}{(m_{\Lambda_{b}}-m_{\Lambda^{*}_{-}})}\Big[\Pi^{T5,B}_{5}+(m_{\Lambda_{b}}-m_{\Lambda^{*}_{+}})\Pi^{T5,B}_{6}\Big]\ ,\nonumber\\
    g^{T5}_{g-}(q^2)&&=0\ .
\end{eqnarray}
where the $\Pi^{d,B}_{i}$ are the coefficients of each Lorentz structure  on the partonic representation after the quark-hadron duality and the Broel transformation by using Eq.(\ref{equ::Borel transformation}).

\section{The correlation matrix of the $z$-series expansion coefficients $a^f_{1,2}$}
\label{sec:Appendix-D}
\begin{sidewaystable}
\caption{ The correlation coefficient matrix between the $z$-series expansion coefficients $a^f_{1,2}$}
    \label{Table:the-correlation-matrix}
    \centering
    \begin{tabular}{|c|ccccccccccccccc|}
    \toprule
    \hline
    \hline
&$a^{f^V_t}_1$&$a^{f^V_0}_0$&$a^{f^V_0}_1$&$a^{f^V_\perp}_0$&$a^{f^V_\perp}_1$&$a^{g^A_t}_1$&$a^{g^A_0}_0$&$a^{g^A_0}_1$&$a^{g^A_\perp}_1$&$a^{f^T_0}_0$&$a^{f^T_0}_1$&$a^{f^T_\perp}_1$&$a^{g^{T5}_0}_1$&$a^{g^{T5}_\perp}_0$&$a^{g^{T5}_\perp}_1$\\
\midrule
\hline
$a^{f^V_t}_1$&1.000&-0.715&0.636&-0.700&0.626&0.620&-0.720&0.622&0.660&-0.700&0.625&0.627&0.660&-0.719&0.622\\
$a^{f^V_0}_0$&-0.715&1.000&-0.968&0.883&-0.786&-0.777&0.905&-0.774&-0.824&0.883&-0.785&-0.786&-0.825&0.905&-0.776\\
$a^{f^V_0}_1$&0.636&-0.968&1.000&-0.778&0.694&0.687&-0.799&0.686&0.729&-0.778&0.693&0.695&0.730&0.798&0.687\\
$a^{f^V_\perp}_0$&-0.700&0.883&-0.778&1.000&-0.971&-0.774&0.901&-0.770&-0.820&0.879&-0.782&-0.782&-0.821&0.901&-0.772\\
$a^{f^V_\perp}_1$&0.626&-0.786&0.694&-0.971&1.000&0.690&-0.802&0.689&0.732&-0.782&0.697&0.698&0.733&-0.802&0.690\\
$a^{g^A_t}_1$&0.620&-0.777&0.687&-0.774&0.690&1.000&-0.800&0.683&0.735&-0.773&0.689&0.691&0.725&-0.793&0.683\\
$a^{g^A_0}_0$&-0.720&0.905&-0.799&0.901&-0.802&-0.800&1.000&-0.943&-0.971&0.901&-0.802&-0.803&-0.843&0.924&-0.794\\
$a^{g^A_0}_1$&0.622&-0.774&0.686&-0.770&0.689&0.683&-0.943&1.000&0.984&-0.769&0.688&0.690&0.726&-0.791&0.686\\
$a^{g^A_\perp}_1$&0.660&-0.824&0.729&-0.820&0.732&0.735&-0.971&0.984&1.000&-0.819&0.731&0.733&0.771&-0.841&0.727\\
$a^{f^T_0}_0$&-0.700&0.883&-0.778&0.879&-0.782&-0.773&0.901&-0.769&-0.819&1.000&-0.971&-0.782&-0.820&0.901&-0.771\\
$a^{f^T_0}_1$&0.625&-0.785&0.693&-0.782&0.697&0.689&-0.802&0.688&0.731&-0.971&1.000&0.697&0.732&-0.801&0.689\\
$a^{f^T_\perp}_1$&0.627&-0.7856&0.695&-0.782&0.698&0.691&-0.803&0.690&0.733&-0.782&0.697&1.000&0.744&-0.808&0.691\\
$a^{g^{T5}_0}_1$&0.660&-0.825&0.729&-0.821&0.733&0.725&-0.843&0.726&0.771&-0.820&0.732&0.744&1.000&-0.972&0.985\\
$a^{g^{T5}_\perp}_0$&-0.719&0.905&-0.798&0.901&-0.802&-0.793&0.924&-0.791&-0.841&0.901&-0.801&-0.808&-0.972&1.000&-0.945\\
$a^{g^{T5}_\perp}_1$&0.622&-0.776&0.687&-0.772&0.690&0.683&-0.794&0.686&0.727&-0.771&0.689&0.691&0.985&-0.945&1.000\\
\hline
\hline
\bottomrule
    \end{tabular}
\end{sidewaystable}

\end{document}